\documentclass{aa}  

\usepackage{graphicx}

\usepackage{txfonts}

\newcommand{\argmin}[1]{\underset{#1}{\operatorname{arg}\,\operatorname{min}}\;}
\newcommand{\argmax}[1]{\underset{#1}{\operatorname{arg}\,\operatorname{max}}\;}

\newcommand{\vl}{\mbox{\boldmath$l$}}
\newcommand{\vm}{\mbox{\boldmath$m$}}

\newcommand{\vw}{\mbox{\boldmath$w$}}
\newcommand{\vx}{\mbox{\boldmath$x$}}

\newcommand{\vf}{\mbox{\boldmath$f$}}
\newcommand{\vy}{\mbox{\boldmath$y$}}
\newcommand{\vu}{\mbox{\boldmath$u$}}
\newcommand{\vn}{\mbox{\boldmath$n$}}
\newcommand{\vs}{\mbox{\boldmath$s$}}

\usepackage{amsmath,amssymb,amsfonts}
\usepackage{algorithmic}
\usepackage{subfig}
\usepackage{graphicx}
\usepackage{float}
\usepackage{textcomp}
\usepackage{xcolor}
\usepackage{graphicx}
\usepackage[ruled,vlined]{algorithm2e}
\usepackage{amssymb}
\usepackage{mathabx}
\usepackage{mathrsfs}
\usepackage{tikz}
\usepackage{pgfplots} 
\usepackage{multirow}

\usepackage{natbib,twoopt}
\usepackage[breaklinks=true]{hyperref} 

\bibpunct{(}{)}{;}{a}{}{,}             
\makeatletter
  \newcommandtwoopt{\citeads}[3][][]{\href{http://adsabs.harvard.edu/abs/#3}%
    {\def\hyper@linkstart##1##2{}%
     \let\hyper@linkend\@empty\citealp[#1][#2]{#3}}}
  \newcommandtwoopt{\citepads}[3][][]{\href{http://adsabs.harvard.edu/abs/#3}%
    {\def\hyper@linkstart##1##2{}%
     \let\hyper@linkend\@empty\citep[#1][#2]{#3}}}
      \newcommandtwoopt{\citeyearads}[3][][]%
    {\href{http://adsabs.harvard.edu/abs/#3}
    {\def\hyper@linkstart##1##2{}%
    \let\hyper@linkend\@empty\citeyear[#1][#2]{#3}}}
\makeatother

\begin{document} 

\title{  Description of turbulent dynamics in the interstellar medium: Multifractal microcanonical analysis:}
\subtitle{II. Sparse filtering of {\sl Herschel} observation maps and visualization of filamentary structures at different length scales. }
\author{ A. Rashidi\inst{1}
 \and H. Yahia\inst{1}
   \and S. Bontemps\inst{2} 
    \and  N. Schneider\inst{3} 
 \and L. Bonne\inst{2}  
  \and P. Hennebelle\inst{4}  \and J. Scholtys\inst{4}
 \and G. Attuel\inst{1}   \and A. Turiel\inst{5}  \and R. Simon\inst{3}    \and   A. Cailly \inst{6} \and A. Zebadua\inst{1} \and A. Cherif \inst{6}  \and C. Lacroix \inst{6} \and M. Martin\inst{1} \and A. El Aouni\inst{1} \and C. Sakka\inst{1} \and S. K. Maji\inst{7}}

\institute{INRIA, Geostat team, 200 rue de la Vieille Tour, 33405 Talence, Cedex, France  
\email{arasheri@gmail.com} 
\and Laboratoire d'Astrophysique de Bordeaux, CNRS UMR 5804,  All. Geoffroy Saint-Hilaire, 33600 Pessac, France
\and I. Physik. Institut, University of Cologne, Z\"ulpicher Str. 77, 50937 Cologne, Germany
\and  Université Paris-Saclay, Université Paris Cité, CEA, CNRS, AIM, 91191, Gif-sur-Yvette, France
\and ICM CSIC, Pg. Marítim de la Barceloneta, 37, Ciutat Vella, 08003 Barcelona, Spain
\and I2S (Innovative Imaging Solutions), 28-30 rue Jean Perrin, 33608 Pessac Cedex, France 
\and IIT Patna, GVP2+7F7, Bihta, Dayalpur Daulatpur, Patna, Bihar, 801106 India
}

\date{Draft of \today}

\abstract{ We present significant improvements to our previous work on noise reduction in {\sl Herschel} observation maps by defining sparse filtering tools capable of handling, in a unified formalism, a significantly improved noise reduction as well as a deconvolution in order to reduce effects introduced by the limited instrumental response (beam). We implement greater flexibility by allowing a wider choice of parsimonious priors in the noise-reduction process. More precisely, we introduce a sparse filtering and deconvolution approach approach of type $l^2$-$l^p$, with $p > 0$ variable and apply it to a  larger set  of molecular clouds using {\sl Herschel}  250 $\mu $m data in order to demonstrate their wide range of application. In the {\sl Herschel} data, we are able to use this approach to highlight extremely fine filamentary structures and obtain singularity spectra that tend to show a significantly less $\log$-normal behavior and a filamentary nature in the less dense regions. We also use high-resolution adaptive magneto-hydrodynamic simulation data to assess the quality of deconvolution in such a simulated beaming framework.}
\maketitle

\section{Introduction}
\label{sec:intro}
\begin{figure*}[h]
	\includegraphics[width=0.485\textwidth]{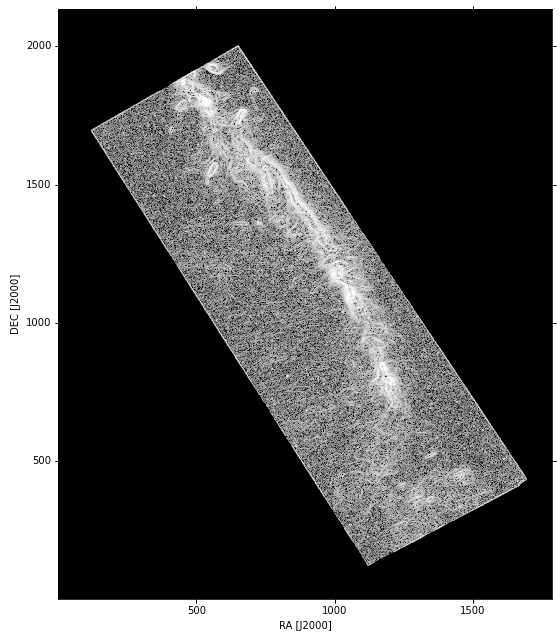}  \includegraphics[width=0.565\textwidth]{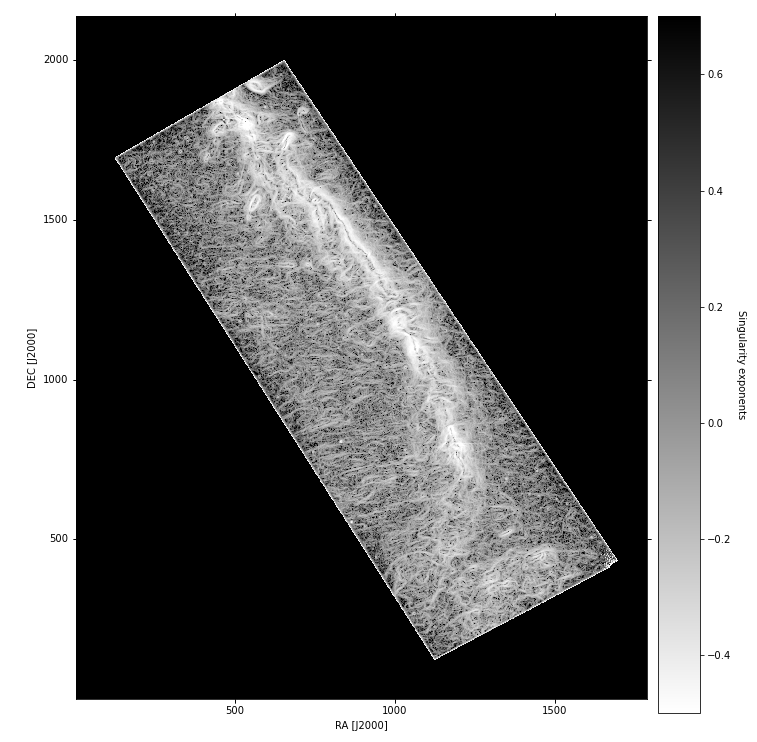}  
	\caption{Singularity exponents of the  {\sl Herschel} observation of Musca at 250 $\mu$m. Left:  After noise reduction using the $l^1$-$l^1$ algorithm presented in~\citep{yahia2021}. Right: After noise reduction, $p=1.5$, $q=-1$, $\lambda=0.1$.}
	\label{musca-se}
\end{figure*}
\begin{figure}[h]
	\includegraphics[width=0.5\textwidth]{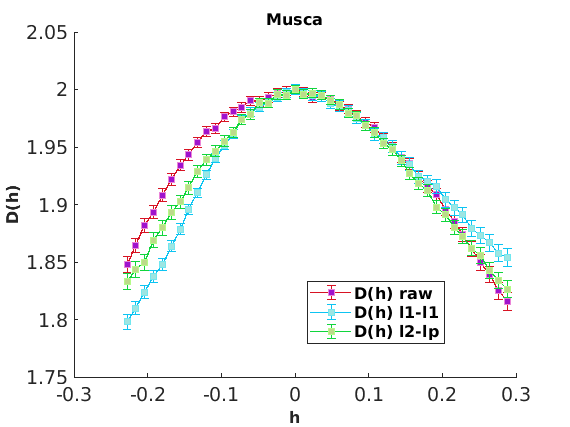} 
	\caption{Comparison of the resulting singularity spectra associated with Fig.~\ref{musca-se}.}
	\label{musca-se2}
\end{figure}
\begin{figure*}[h]
	\centering
	\includegraphics[width=1.0\textwidth]{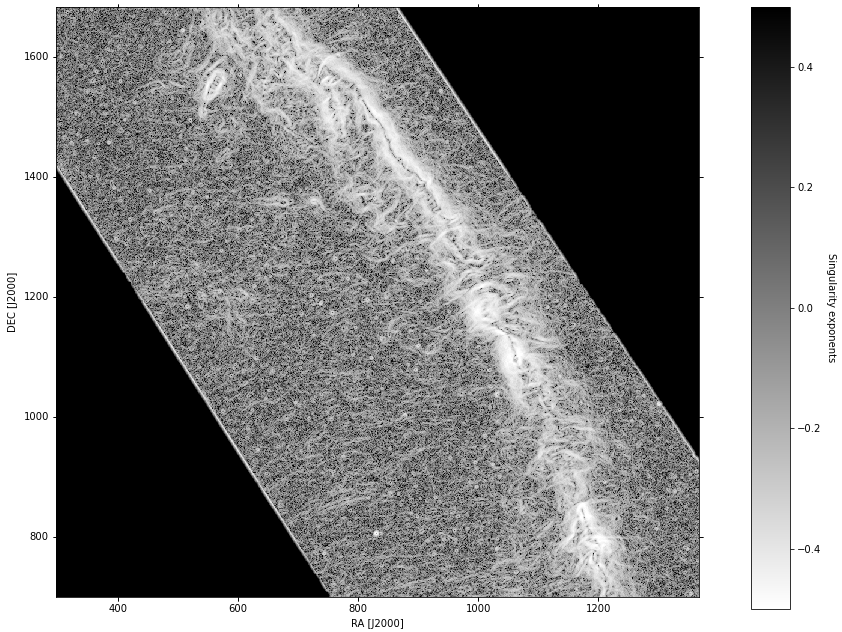} 
	\caption{Singularity exponents of the {\sl Herschel} observation of Musca at 250 $\mu$m. Here we show a zoom onto a particular subregion. The noise reduction with the $l^1$-$l^1$ algorithm corresponds to Fig. 7 of~\citet{yahia2021}.}
	\label{musca-se-zooma}
\end{figure*}
\begin{figure*}[h]
	\centering
	\includegraphics[width=1.0\textwidth]{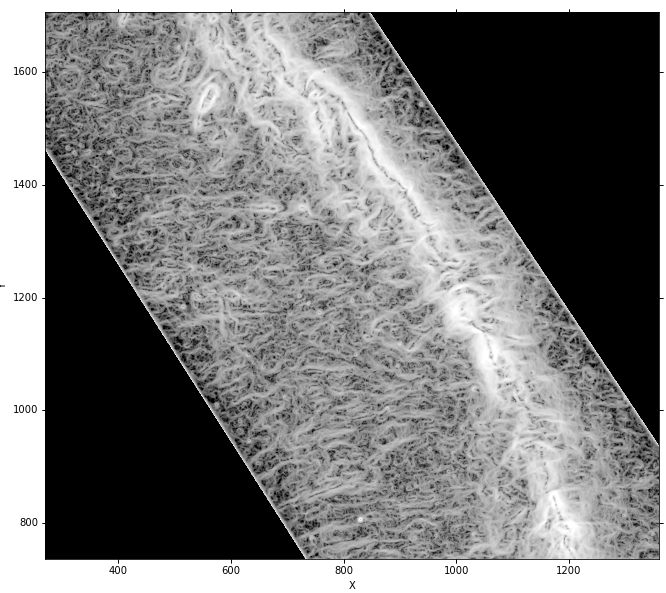} 
	\caption{Singularity exponents of the {\sl Herschel} observation of Musca at 250 $\mu$m. Here we show a zoom onto a particular subregion:  $l^2$-$l^p$ noise reduction with  $p=1.5$, $q=-1$, $\lambda=0.1$. The new noise-reduction algorithm exposes more filamentary structures. Here we show the same gray-level map as that in Fig.~\ref{musca-se-zooma}.}
	\label{musca-se-zoomb}
\end{figure*}
\paragraph{}{Observations with the {\sl Herschel} space observatory have highlighted the ubiquity of filamentary structures at different spatial scales in the molecular clouds of the Milky way~\citep{Arzoumanian2011,Hill2011,Schneider2012,Schisano2014,Cox2016}, which raises the question of their supposed role in the star formation process (the literature on this subject is vast; see for instance~\citep{Andre2014,2022arXiv220309562H}. Numerous studies of the dynamical and geometrical structure of filaments on cloud scales (from subparsec scales to  scales of a few parsecs) reveal that gas accretion onto filaments and subsequent fragmentation set the initial conditions of star formation. However, the relative importance of turbulence, external pressure, magnetic fields, and gravity in filament formation and evolution is still being explored and strongly debated. There have been a number of proposals for how filaments, or more general dense structures, form. One scenario proposes that magnetized HI bubbles (triggered by supernova explosions) could trigger compression, which initiates mass accumulation along the magnetic field lines on the dense filaments~(\citep{Inoue2009,Inoue2018, Bonne2020a,Bonne2020b,Schneider2023}). This could fit with the observed presence of large bubbles in the solar neighborhood~\citep[e.g.,][]{zucker2023solar,pinedaproc2023}. However, other simulations explain cloud formation with colliding large-scale flows of HI, where dense gas, often in filamentary structures, forms in shock-compressed layers \citep{Koyama2000,Clarke2019,Dobbs2020}. In parallel to this large-scale (over tens of parsecs) assembly and structuring of clouds, material flows via fainter filaments (striations) on smaller scales (parsecs) onto larger filaments. The densest ones are called ridges~\cite{Schneider2010,Hennemann2012}. In addition to converging flows as a possible mechanism for the formation of HI clouds at large scales ($>$10pc),  there is vast literature on galaxy-scale mechanisms, such as disk instabilities~\citep{KimOstriker2001,Dobbsetal2008} and gas-arm interactions~\citep{Bonnelletal2006}. New essential questions are therefore related to  whether the different hierarchial levels are interconnected and whether or not certain physical processes dominate at the different scales. In line or continuum maps of any cloud, we observe this hierarchy and our objective is to provide measurements that can be used  to quantify spatial structures and to link them to the underlying physical processes.}
\paragraph*{}{\citet{yahia2021} and \citet{Robitaille2019} showed that to accurately access the multiscale properties of a {\sl Herschel} observation map,  it is necessary to  work on a reduction of the noise present in the {\sl Herschel}  data. Such noise comes from the cosmic infrared background (CIB), the cosmic microwave background (CMB), and from the instrument itself, among other disturbances. Although the noise of the instrument may be of a fractal nature, the most important components of the noise come from CMB and CIB.  The noise from the former can be seen as a Gaussian process to first approximation~\citep{Buchert_2017}, and tends to modify the shape of  a singularity spectrum computed on an observation  map, making it more symmetrical, thus concealing the statistics coming from the filamentary structures at small scales, and consequently misleading interpretation of a singularity spectrum in terms of the underlying dynamics. Indeed, a Gaussian process may be mono- or multifractal, and stationary or not; but in all cases it is characterized by second-order statistics only. In addition, the reduction of Gaussian noise makes it possible to visualize ---via the geometric distribution of singularity exponents on an observation map--- the extraordinary complexity of the spatial distribution of filamentary structures at small scales; that is, structures that are responsible for the non-$\log$-normal character of the obtained singularity spectra. In~\citet{yahia2021}, we solved the problem of noise reduction by considering a sparse noise-reduction algorithm that tends to eliminate Gaussian noise while keeping the coherent gradient information at small scales. This was done by solving the following $l^1$ optimization problem:
	\begin{equation}
		\centering
		\underset{\vs_f}{\operatorname{argmin}} ~~~ \| \vs - \vs_f \|_1 + \lambda \| \nabla \vs_f \|_1 
		\label{eqn:l1-opt}
		,\end{equation}
where $\vs$ is the original (noisy) observation map,  $\vs_f$ is the filtered (with noise reduction) computed map, and $\lambda > 0$ is a tuning parameter. The first term $ \|  \vs-  \vs_f \|_1$ guarantees that the filtered image is very similar to the original (data fitting term), while the second term (also known as {\sl prior}), $ \lambda \,\| \nabla \vs_f \|_1$, when minimized, promotes the sparsity of the filtered gradients~\citep{8187544}. Different data-fitting terms can be introduced depending on the type of noise present in the data (Gaussian, Laplace, Poisson, etc.)~\citep{Getreuer2012}. This type of approach to the problem of noise reduction is a generalization of the total variation (TV)\footnote{The total variation is the integral of the gradient norm. Noisy signals tend to display a high TV value.} methods~\citep{5413587} which have taken on great importance in image processing since the founding article~\citep{RUDIN1992}. In particular, sparse $l^1$ and the TV scheme are applied in The Event Horizon Telescope Collaboration~\citep{EHT4-2019}.}
\paragraph*{}{Here we present improvements to the Gaussian noise-reduction problem by considering other types of priors, which are more parameterizable and allow finer control over the preservation of filamentary structures. We demonstrate the spectacular improvements in visualization of these filamentary structures at different scales, thus exposing the complexity of the geometric organization of the interstellar medium (ISM) as accessible in the {\sl Herschel} data, and study the quality of the singularity spectra obtained on filtered and debeamed data. Moreover, the filtering method taken in this study also includes beam reduction in a blind deconvolution formulation, which contributes, in a unified sparse approach, to both noise and beam reduction.}
\paragraph*{}{We begin by introducing the general model of deconvolution and noise reduction in Sect.~\ref{sec:sparse}.  In Sect.~\ref{sec:processing}, we then reiterate the differences in physical processes signed by singularity spectra, namely  $\log$-normal and $\log$-Poisson, 
from our previous work~\citep{yahia2021} , and we demonstrate the superiority of our new algorithm over the Musca observation map.  In section~\ref{sec:deconvolution}, we present the magnetohydrodynamic simulation data used in this work, which serve as "ground truth" to evaluate the reduction of the beam effect and the deconvolution algorithm.  Section~\ref{sec:hdata} presents the results obtained on several {\sl Herschel} observation maps;  here we also show how the tools developed in this study make it possible to separate ---in the {\sl Herschel} data that we studied--- the turbulence properties of dense regions containing protostars from the turbulence properties of regions that are more diffuse and filamentary. We end with a discussion of our results and the conclusions that we draw from this work.}

\section{Sparse approach to noise reduction and deconvolution}
\label{sec:sparse}
\paragraph{}{In this work, we focus on a formulation of noise reduction and debeaming as an inverse problem~\citep{Hansen2010,Dupe2012,OPT-003,Bertero2021}. Inverse problems arise naturally in many areas of signal and image processing.  In the present case, we consider an observational map \vs\  containing information perturbed by noise and beam effects; the original or true observation map is the ideal representation of the observed scene. Generally, the observation process is never perfect: there is uncertainty in the measurements, occurring as beam convolution, noise, and other degradation in the recorded observation map. The aim of digital image restoration is to recover an estimate of the original observational map from the degraded observations. The key to being able to solve this {ill-posed} inverse problem is proper incorporation of prior knowledge about the original observation map into the restoration process~\citep{badri:tel-01239958}. The many existing studies on interstellar clouds describe them as containing filamentary structures at different scales, which leads to favoring priors respecting sparse gradients.}
\paragraph{}{Image deconvolution is the technique of computing a sharper reconstruction of a digital image from a blurred and noisy one based on a mathematical model of the blurring process (beam plus noise). In astrophysics, noise reduction and beam deconvolution must be performed without notable deterioration in the statistics of the dynamical system observed, which is a very challenging requirement. Image deconvolution is mainly divided into two categories: blind and nonblind deconvolution~\citep{YIN202113}.  Nonblind image deconvolution seeks an estimate of the true observational map assuming the blur is known.  In contrast, blind image restoration tackles the much more difficult (but realistic) problem where the degradation is unknown.  In our problem, we are dealing with nonblind image deblurring (NBID) where the observed image is modeled as the convolution of an underlying (sharp) image and a known blurring filter (the instrumental beam), often followed by additive noise.}
\paragraph*{}{The general, discrete model for linear degradation caused by beam and additive noise is formulated as:
	\begin{equation}        
		\vy(\vx) = \displaystyle \sum_{\vw \in {\cal S}_{\mathbb{H}}}\mathbb{H}(\vx,\vw)\vs(\vw)+\vn(\vx) 
		\label{model:noise}
		,\end{equation}
	where $\vx=(x,y)\in \Omega$ is a spatial location in $\Omega$, the domain of the observation map $\vs$; $\vy(\vx)$ is the observed image;  $\mathbb{H}(\vx,\vw)$ is the  point spread function (PSF); $\vn(\vx) $ is the noise; and ${\cal S}_{\mathbb{H}} \subset  \mathbb{R}^{2}$ is the support of the PSF. The additive noise process $\vn(\vx) $  originates during the acquisition of the observation map. Common types of noise are electronic, photoelectric, quantization noise, and ---most important in our case--- noise coming from CIB, CMB, far distant background objects, and so on.
	The image-degradation model described by equation~\ref{model:noise} is very commonly represented in terms of a matrix-vector formulation:
	\begin{equation}
		\vy={\bf H}\vs+\vn
		\label{model:noise-digital}
		,\end{equation}
	where $\bf H$  is a matrix representation of a convolution operator $\mathbb{H}$  (PSF); if this convolution is periodic, $\bf H$ is then a (block) circulant matrix. In the present paper, the beam effect in the images is modeled in the matrix $\bf H$.  The goal is to recover $\vs$ from the observational map $\vy$.  In the inverse problem of equation~\ref{model:noise}, which is ill-conditioned, the maximum likelihood estimation (MLE) $\hat{\vs}$ of $\vs$  is defined by:
	\begin{equation}
		\hat{\vs} = \hat{\vs}(\vy) = \argmax{\vs} {\bf p}_{\vs|\vy}(\vs|\vy)
		\label{eqn:MAP1}
		,\end{equation}
	where ${\bf p}_{\vs|\vy}(\vs|\vy)$ is the conditional probability distribution of $\vs$ knowing $\vy$. From this, we obtain\begin{equation}
		\begin{array}{lcl}
			\hat{\vs}  & = &        \argmax{\vs} {\bf p}_{\vy|\vs}(\vy|\vs) {\bf p}_{\vs}(\vs) \\
			~ & = &         \argmax{\vs} {\bf p}_{\vn} (\vy - {\bf H}\vs) {\bf p}_{\vs}(\vs)
		\end{array}
		\label{eqn:MAPDERIVATION}
		.\end{equation}
	Taking logarithms in eq.~\ref{eqn:MAPDERIVATION} leads to
	\begin{equation}
		\hat{\vs} =  \argmax{\vs} \log {\bf p}_{\vn} (\vy - {\bf H}\vs) + \log {\bf p}_{\vs}(\vs)
		\label{eqn:MAP2}
		,\end{equation}
	where ${\bf p}_{\vn}$ and ${\bf p}_{\vs}$ denote the probability laws of the noise $\vn$ and random variable $\vs,$ respectively. When some assumption is made on ${\bf p}_{\vs}$, the MLE is called maximum a posteriori (MAP) estimation, and the MAP estimate takes the form of an optimization problem:
	\begin{equation}
		\hat{\vs} =  \argmin{\vs} \vf(\vs,\vy).
		\label{eqn:argmingeneral}
	\end{equation}
	In \textcolor[rgb]{0,0,0}{what follows,} we assume that $\vn$ is Gaussian, so that $\vf(\vs,\vy)$ can be decomposed as a sum of two terms, the first one being an $L^2$ norm and the MAP estimate of $\vs$ given $\vy$ is written as~\citep{lorenz2007}:
	\begin{equation}
		\vf(\vs,\vy) = \| {\bf H} \vs-\vy \|^2+\lambda \phi({\cal D }\vs).
		\label{eqn:math}
	\end{equation}
	In this equation, the first term, $ \| {\bf H} \vs-\vy \|^2 $, is the data fidelity term. The second term,  $\lambda \phi$, is a regularization function that enforces prior knowledge about $\vs$ into the solution. We note that $\phi$ corresponds to the entropy of the prior $\log {\bf p}_{\vs}(\vs)$. Depending on the assumption of the characteristic of the image or observational map, various $\phi$ can be chosen in the cost function. It must be noted that, for some interesting priors, $\phi$ may be neither differentiable nor convex; for example, choosing $\phi(\vs) = \|\vs\|_p^p$ with $0 < p < 1$ ($l^p$ quasi-norm)  defines a nonconvex prior. Consequently, one must use the most advanced optimization techniques in the numerical implementation of eq.~\ref{eqn:argmingeneral}; for instance, the ones using the notion of a proximal operator, as described in  Appendix~\ref{sec:sol}. Our observation of images is carried out in the gradient domain, which means that if ${\cal D}$ denotes the discrete gradient operator, for simplicity we write $ \phi({\cal D }\vs)$ instead of $ \phi({\cal D }_x\vs)+ \phi({\cal D }_y\vs),$  which are respectively the matrix form of the first-order gradients in both directions:  $d_x=[1, -1]$ and $d_y=[1, -1]^\top$.}
\paragraph*{}{In the remainder of this work, ${\bf H}$ is the PSF of the beam effect, and the choice of the function $\phi$ is explained by equation Eq.~\ref{definition-potential} of Appendix~\ref{sec:reg}. Therefore, we solve the problem of deconvolution and noise reduction by minimizing the functional of Eq.~\ref{eqn:math}, in which {\bf H} and $\vy(\vx)$ are the data of the problem, and the solution is the calculated minimum $\hat{\vs} $. When we set ${\bf H} = \mbox{\bf Id}$ (identity matrix), then there is no deconvolution and the minimization of the functional of equation eq.~\ref{eqn:math} corresponds to noise reduction only. Concerning the choice of the function $\phi$, which is defined in Appendix B via potential functions $\varphi_q^p$ applied to each coordinate:
	\begin{equation*}
		\phi(\vu) = \displaystyle \sum_i \varphi_q^p(u_i),
	\end{equation*}
	the chosen values of $p$ and $q$ are detailed in Appendix~\ref{sec:choice}. In most cases, $q=-1$ and $p$ is in the interval $[1.3, 1.7],$ which corresponds to the choice of a $l^p$ norm:
	$\phi(\vu)=||\vu||^p_p$ in the functional definition in Eq.~\ref{eqn:math}. When this is the case, eq.~\ref{eqn:math} defines two terms, the first being a $l^2$ norm and the second a $l^p$ norm, hence the name {\it $l^2$-$l^p$} deconvolution algorithm in that case; this can be compared with eq.~\ref{eqn:l1-opt},  which is of {\it $l^1$-$l^1$} type because of the presence of two $l^1$ norms in its formulation. We show the optimization method used to solve the problem in eq.~\ref{eqn:math} in Appendix A.}
\section{Analysis of an observation map}
\label{sec:processing}
\paragraph*{}{The results described in this study make use of the methodology detailed in~\citet{yahia2021}, to which we refer the reader for further details.}
 \subsection{The $\log$-Poisson process}
 \paragraph*{}{Among the results obtained in our previous study, in particular in regards to the Musca observation map, is the finding of nonsymmetrical singularity spectra, and in particular nonparabolic spectra. We reiterate here that a $\log$-normal process in $\mathbb{R}^d$ has a parabolic singularity spectrum given by the formula:
 	\begin{equation}
 		\label{lognormalspectrum}
 		D({\bf h}) = \displaystyle  d- \frac{1}{2} \left ( \frac{{\bf h} - {\bf h}_m}{\sigma_{{\bf h}}} \right )^2,
 	\end{equation} 
 where ${\bf h}_m$ and $\sigma_{{\bf h}}$ are the average and dispersion of the singularity exponents, respectively (see~\citep{yahia2021} for details). Each time one observes a nonsymmetrical singularity spectrum on a real observation map, they are faced with the problem of the physical interpretation ---or explanation--- of this result. As mentioned in our previous study, a $\log$-normal process has a simple physical interpretation in terms of a multiplicative cascade of a particular type, as commonly invoked in many astrophysics articles and studies dealing with the turbulence of the interstellar medium~{\citep{1Tassis2010,bron:tel-01111148,Corbelli2018,Mocz2019,Mattsson2020,Bellomi2020}}. In contrast, in regards to turbulence,  there exists a multiplicative cascade-type phenomenology involving $\log$-Poisson processes: the transfer of energy between scales is carried out via multifractal geometric structures much more complex than those corresponding to the $\log$-normal model~\citep{PhysRevLett.72.336}. Regarding turbulence in the ISM, although it may be logical to invoke this type of the non$\log$-normal multiplicative cascade, we believe that this needs to be substantiated by observations, because there are many processes other than $\log$-Poisson that have an asymmetric spectrum (e.g., $\log$-$\alpha$-stable processes~\citep{Renosh2015}); notwithstanding that there is not yet any physical justification for  the majority of these processes. Moreover, with regards to the non$\log$-normal processes present in ISM turbulence, the  role of specific physical phenomena (magnetic field, stellar feedback, etc.) with respect to the filamentary structures requires an in-depth study supported by observations.}
\paragraph*{}{The left panel of Figure~\ref{musca-se}  first shows the map of the singularity exponents of the $l^1$-$l^1$ filtered observation map, and  the right panel shows the resulting singularity exponents after having applied the noise-reduction procedure explained in section~\ref{sec:sparse} with $p=1.5$, $q=-1$, and $\lambda=0.1$. Figure~\ref{musca-se2} shows three singula\-rity spectra: that of the raw unfiltered {\sl Herschel} data, the $l^1$-$l^1$ filtering result, and the filtering result with $p = 1.5$, $q=-1,$ and $\lambda = 0.1,$ which corresponds to $l^2$-$l^p$ with $p=1.5$. The comparison of the spectra on the most singular part, that is, the part corresponding to ${\bf h} < 0,$ illustrates the type of trade-off that can arise with this noise reduction approach by energy minimization: decreasing $p$ favors the most singular structures, that is, the thinnest filaments, but at the expense of noise elimination, and therefore of their visualization. On the other hand, increasing $p$ improves noise reduction, and therefore the visualization of filaments, but at the expense of a "correct" evaluation of the fractal dimension of these components. This is because the more $p$ increases, the more the filaments tend to thicken. There is no "miracle solution" to this type of problem; hence the trade-off. Different values close to $p$ can be chosen either for the visualization of the filaments hidden in the noise, or for the most exact calculation possible of the singularity spectrum. We note that the spectrum $l^2$-$l^p$ nevertheless remains below that of the raw data for the values of $p$ reinforcing the visualization of the filaments, which is the desired outcome. Figures~\ref{musca-se-zooma} and \ref{musca-se-zoomb} show the map of the singularity exponents of a subregion of the main filament with both algorithms; the extreme complexity of linear structures at lower scale is now beco\-ming even more evident.}

\section{Experiments on magnetohydrodynamic simulation data}
\label{sec:deconvolution}
\paragraph*{}{In this section, we study the results of the proposed approach applied to magnetohydrodynamic simulation data. Our objective  in using these data is twofold:
	\begin{enumerate}
		\item  To study the elimination of the beam effect using our approach on data with a "ground truth".
		\item  To present high-spatial-resolution MHD data that will be explored further in future studies.
\end{enumerate}}
\begin{figure*}[h]
	\centering
	\includegraphics[width=0.32\textwidth]{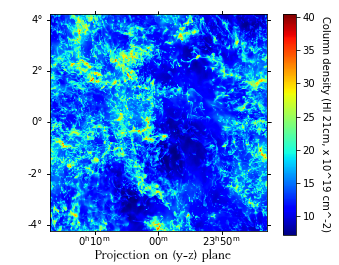}  \includegraphics[width=0.32\textwidth]{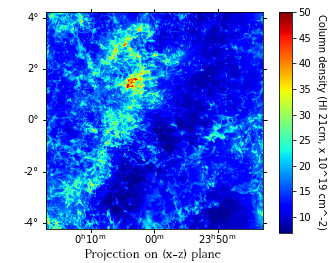}  \includegraphics[width=0.33\textwidth]{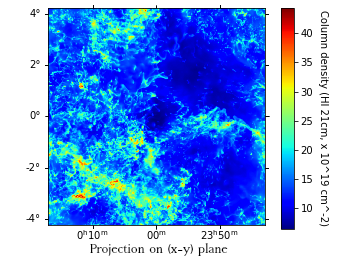} 
	\caption{Simulation output fileA in the $x$, $y,$ and $z$ views (see Table~\ref{tbl:filenames} for an explanation of the filename syntax).}
	\label{MHD-sim1}
\end{figure*}   
\begin{figure*}[h]
	\centering
	\includegraphics[width=0.3\textwidth]{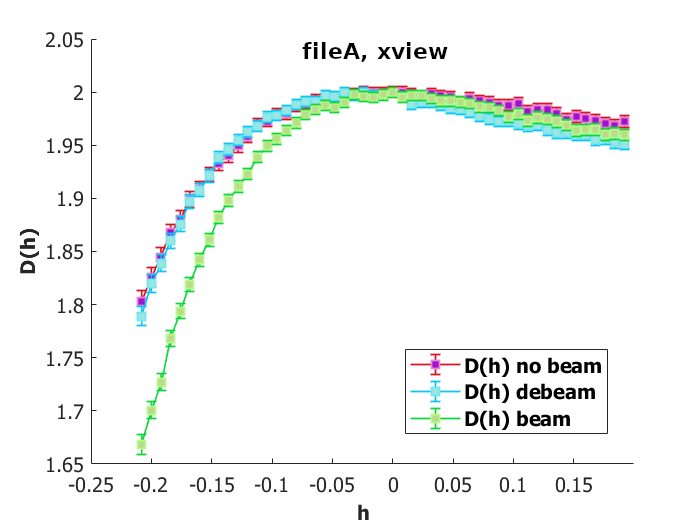}  \includegraphics[width=0.3\textwidth]{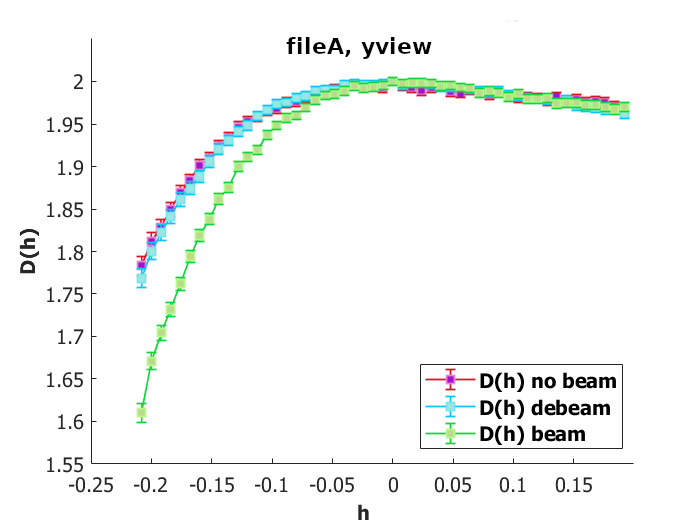}  \includegraphics[width=0.3\textwidth]{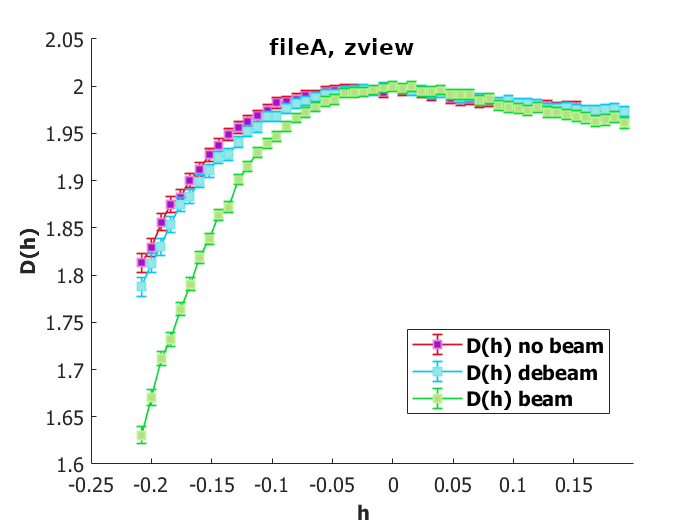} 
	\caption{ Beam reduction performed on simulation output fileA (see Table~\ref{tbl:filenames}). In each image panel we show: the singularity spectrum of the data generated without any beam effect (called "no beam", in red, and corresponding to a simulated "ground truth"),  the singularity spectrum of the data generated with beam effect as described in the main text (2D Gaussian kernel of FWHM = 2 pixels, called "beam" in the graph, in green), and the singularity spectrum of the data after application of the proposed deconvolution algorithm (called "debeam", in light blue). The left image panel shows the $x$ view, filtered with  $p=1.7$ and $\lambda = 0.1$; the middle image panel displays the $y$ view, filtered with  $p=1.5$ and $\lambda = 0.1;$ and the right panel shows the $z$ view, filtered with  $p=1.4$ and $\lambda = 0.1$.}
	\label{spectra-MHD1}
\end{figure*}
\begin{figure*}[h]
	\centering
	\includegraphics[width=0.32\textwidth]{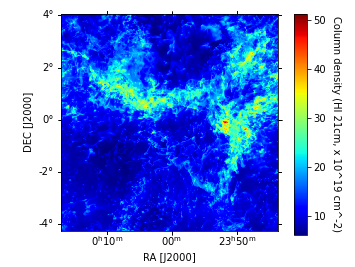}  \includegraphics[width=0.32\textwidth]{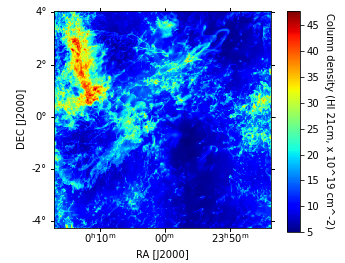}  \includegraphics[width=0.33\textwidth]{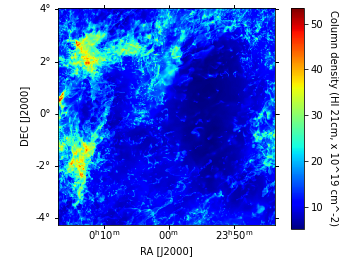} 
	\caption{Simulation outputs fileB in the $x$, $y,$ and $z$ views (see Table~\ref{tbl:filenames} for an explanation of the filename syntax).}
	\label{MHD-sim1B}
\end{figure*}   
\begin{figure*}[h]
	\centering
	\includegraphics[width=0.3\textwidth]{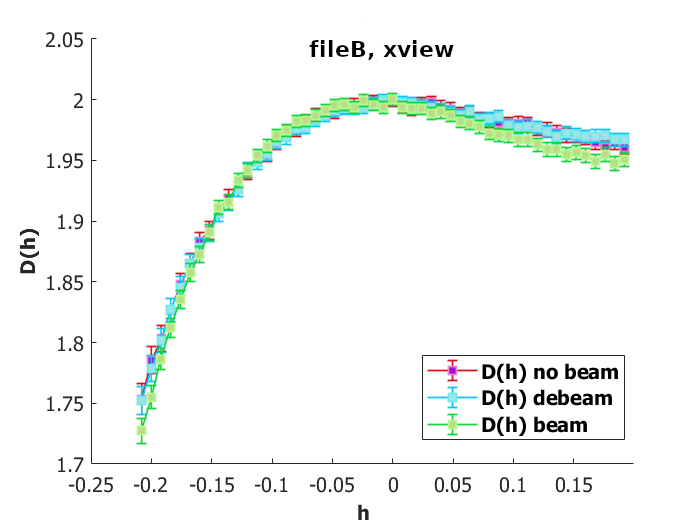}  \includegraphics[width=0.3\textwidth]{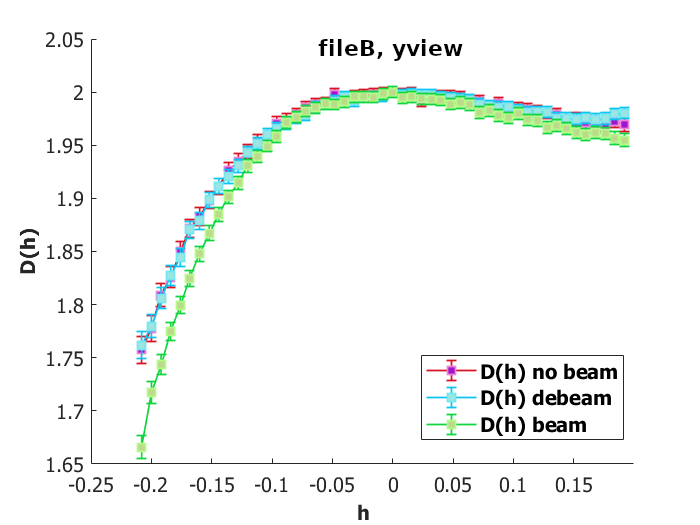}  \includegraphics[width=0.3\textwidth]{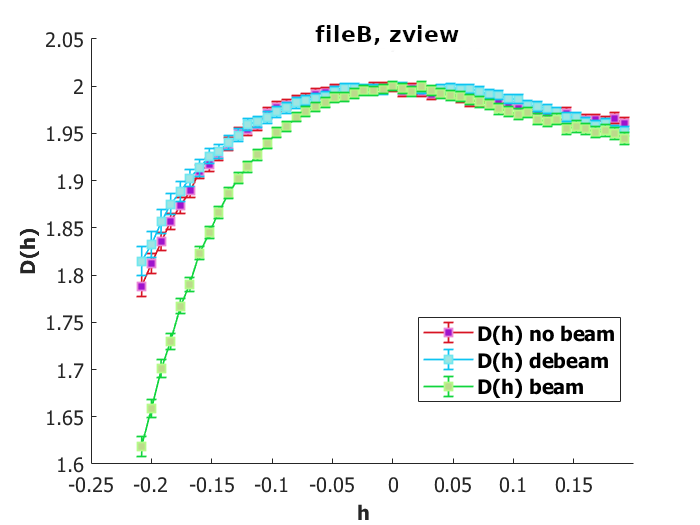} 
	\caption{Beam reduction performed on simulation output fileB (see Table~\ref{tbl:filenames} and Fig.~\ref{spectra-MHD1}). The left image panel shows the $x$ view, filtered with  $p=1.4$ and $\lambda = 0.1$; the middle image panel displays the $y$ view, filtered with  $p=1.4$ and $\lambda = 0.1;$ and the right panel shows the $z$ view, filtered with  $p=1.35$ and $\lambda = 0.1$.}
	\label{spectra-MHD2}
\end{figure*}
\begin{figure*}[h]
	\centering
	\includegraphics[width=0.31\textwidth]{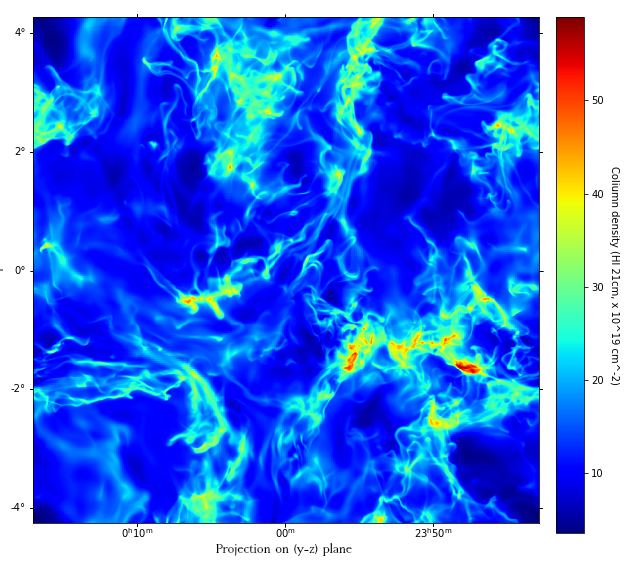}  \includegraphics[width=0.31\textwidth]{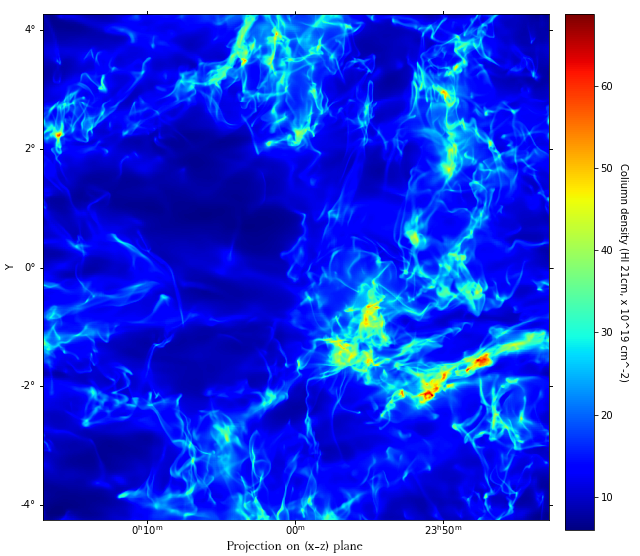}  \includegraphics[width=0.32\textwidth]{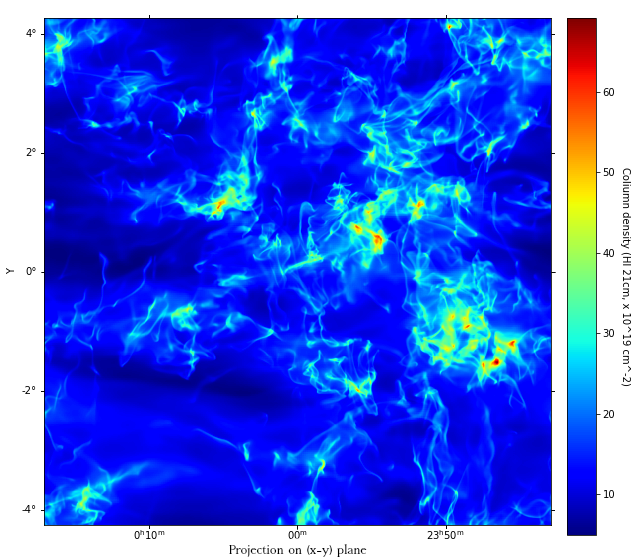} 
	\caption{Simulation outputs fileC in the $x$, $y,$ and $z$ views (see Table~\ref{tbl:filenames} for an explanation of the filename syntax).}
	\label{MHD-sim2}
\end{figure*}
\begin{figure*}[h]
	\centering
	\includegraphics[width=0.32\textwidth]{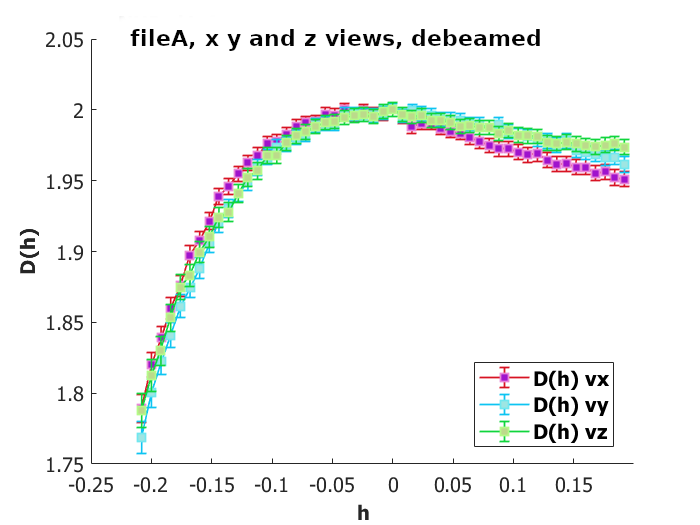} 
	\includegraphics[width=0.32\textwidth]{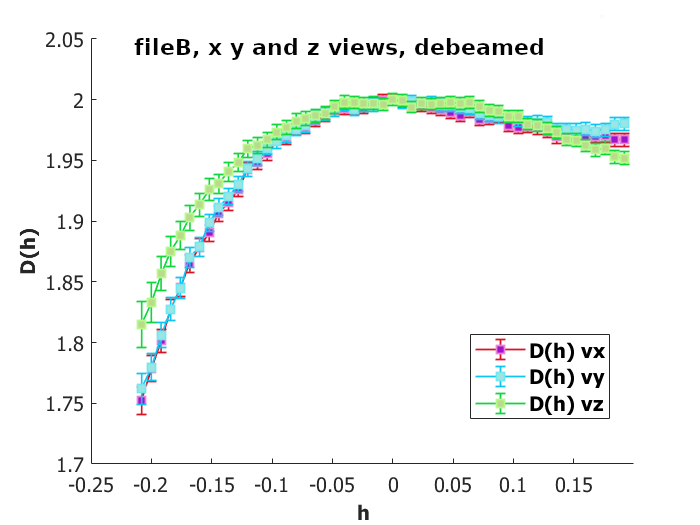} 
	\includegraphics[width=0.32\textwidth]{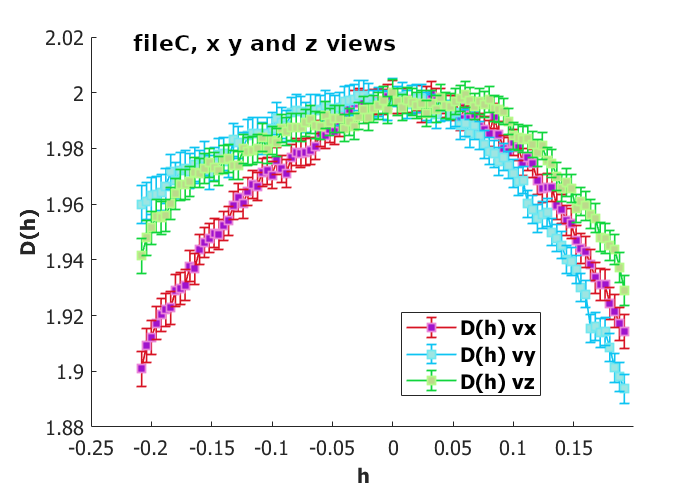} 
	\caption{Comparison between singularity spectra. Left: Comparison of the singularity spectra of  output fileA in the $x$, $y,$ and $z$ views, with  $p=1.7$ for $x$, $p=1.5$ for $y$, $p=1.4$ for $z$ and $\lambda = 0.1$. The three spectra are coincident, indicating same turbulence statistics in the three directions of space. Middle: Comparison of the singularity spectra of  output fileB in the $x$, $y,$ and $z$ views, with  $p=1.4$ for $x$, $p=1.4$ for $y$, $p=1.35$ for $z$ and $\lambda = 0.1$. There is anisotropy in the $z$-direction. Right: Comparison of the singularity spectra of  output fileC in the $x$, $y,$ and $z$ views, with  $p=1.7$ for $x$, $p=1.6$ for $y$, $p=1.6$ for $z$ and $\lambda = 0.1$. There is anisotropy in the $x$-direction.}
	\label{sim1-spectra-x-y-z}
\end{figure*}
\paragraph*{}{Synthetic column density maps of the 21cm hydrogen line were generated from a set of magnetohydrodynamic turbulence simulations of nonisothermal atomic gas designed to statistically reproduce observations of nonself-gravitating cirrus clouds transiting from atomic to molecular gas (Scholtys et al., in prep.). The simulations were carried out with the RAMSES code~\citep{Teyssier2002,Fromang2006} and use a high-order godunov scheme to compute fluxes at the interface between cells. The ideal MHD equations are solved with the HLLD Riemann solver (Miyoshi \& Kuzano 2005). The nullity of $\nabla\cdot \vec{B}$ is guaranteed by the use of the constrained transport method. These simulations leverage the adaptive mesh refinement capabilities of RAMSES. With a coarse grid of 256$^{3}$cells, a cubic volume with side length $L=50\,$pc is refined based on two density thresholds up to effective resolutions of $512^{3}$ and $1024^{3}$. The selected thresholds are $n=10\,$cm$^{-3}$ and $n=20\,$cm$^{-3}$. They were chosen in order to capture the phase transition from warm to cold atomic gas through thermal instability~\citep{Field1965} while maximizing the resolution of the cold structures and enabling a large number of simulations to be performed. The density and temperature dependant cooling function implemented in RAMSES is based on the work by Wolfire et al.~\citep{Wolfire1995,Wolfire2003}. The main heating contribution is from the photoelectric effect on small dust grains and polycyclic aromatic hydrocarbons and the cooling occurs through radiative recombination of electrons onto dust grains in addition to emission lines of $C\mathrm{II}$, O$\mathrm{I,}$ and Lyman$\alpha$ photons following collisional excitation by electrons and hydrogen atoms.}
	
\paragraph*{}{All the simulations start off in typical warm neutral  medium (WNM) conditions with static warm neutral gas with a density of $n=1\,$cm$^{-3}$ and a   temperature of $T=8000\,$K$\,$cm$^{-3}$  with a uniform magnetic field along the $x$ axis $b_x$. A turbulent velocity field is then generated through forcing in Fourier space with a parabolic distribution on wavenumbers $1<k<3$ centered on $k=2$, which is half the size of the box $L/2$. Owing to this wavenumber distribution and to periodic boundary conditions, our simulations therefore model the subvolume of a turbulence cascade generated though energy injection on large scales. The turbulent forcing is continuously applied throughout the evolution of the simulations and takes the form of an acceleration field modeled by an Ornstein-Ulhenbeck stochastic process~ \citep{eswaran-pope1988,schmidt2006,schmidt2009,saury2014}.  The forcing amplitude can be set with the parameter $rms$ and regulates the velocity dispersion amplitude developed by the flow. The percentage of compressible modes or solenoidal modes can be selected through the input $c_f$, where $c_f=0$ and $c_f=1$ mean purely solenoidal and purely compressible modes, respectively. As expected for turbulence driven in compressible gas, shearing motions and shocks will develop within the volume, and so locally the compressible fraction may differ from the large-scale driving applied with the forcing. The simulations are evolved until they reach a stationary state and at least three dynamical time steps $t_{\mathrm{dyn}}=L/\sigma_{3D}$ have elapsed, where $\sigma_{3D}$ represents the 3D turbulent velocity dispersion.}
\paragraph*{}{Two synthetic datasets created from the simulation outputs are 21cm brightness temperature cubes and their corresponding column density maps. As realistic 21cm spectra are required in the analysis done by Scholtys et al. (in prep.), self-absorption of the 21cm line by cold gas in between the observer and the emitting cell is also accounted for; although the synthetic column density maps used here are still integrated from the brightness temperature profiles assuming an optically thin medium. The calculation of the brightness temperature assumes that along a line of sight, every individual cell is at local thermodynamic equilibrium (LTE) and contributes an amount of emission that is a function of its density, temperature, and velocity projection on the line of sight to the overall emission
at velocity $u$. For neutral hydrogen gas at LTE, the thermal velocity distribution is a Maxwell-Boltzmann distribution with thermal velocity dispersion $\sigma_{T}=\sqrt{k_{B}T/m_{H}}$ and the spinning temperature is equal to the kinetic temperature $T$ ($k_B$: Boltzmann constant, $m_H$: mass of hydrogen). For a line of sight along the $z$ axis, the optical depth $\mbox{d}\tau(x,y,z,u)$ at velocity $u$ of a cell of density $n(x,y,z)$, kinetic temperature $T(x,y,z)$, physical size $\mbox{d}z,$ and velocity $v_z(x,y,z)$ is given by:
	\begin{equation}
		\mbox{d}\tau(x,y,z,u)=\frac{n(x,y,z)\,\mbox{d}z}{C \,T(x,y,z)}\frac{e^{-(u-v_z(x,y,z))^{2}/\sigma_{T}^{2}}}{\sqrt{2\pi}\,\sigma_{T}(x,y,z)}.
\end{equation}
The constant $C=1.823\times 10^{18}\,$cm$^{-2}$(K$\,$km$\,$s$^{-1}$)$^{-1}$ is specific to the 21cm line and incorporates the Einstein coefficient for spontaneous emission for this transition and the degeneracy of the associated energy levels. For a line of sight along the $z$ axis, the complete brightness temperature spectra $T_B(x,y,u)$ are obtained through summation over each cell and weighing their emission by the absorption of the foreground cells at $z'<z$:
\begin{equation}
T_B(x,y,u)= \displaystyle \sum_z T(x,y,z) \left(1-e^{-\tau(x,y,z,u)}\right)e^{-\left( \displaystyle \sum_{z' < z}\tau(x,y,z',u) \right)}.\end{equation}
Real observations of a 21cm emission spectrum seldom have an absorption counterpart to estimate the true column density. However, self-absorption corrections are expected to be small over the column density range of the simulations, which is $N_{\mathrm{HI}}\leq 5\times 10^{20}$cm$^{-2}$\citep{Lee_2015,Murray_2018}. Therefore, column density maps are generated under the optically thin approximation, with an integration of the brightness temperature over velocity:
\begin{equation}
		N_{\mathrm{HI}}(x,y)=C \int T_B(x,y,u)\mathrm{d}u.\end{equation}}
\paragraph*{}{The beam matrix applied to the MHD simulation $N_{\mathrm{HI}}$ maps to replicate the instrumental smoothing is a 2D Gaussian kernel of $\mathrm{FWHM}=2\,$pixels, or equivalently $\sigma \simeq 0.849$ pixels, because $\mathrm{FWHM} =  \sqrt{8 \ln 2}\sigma$. The resulting Gaussian kernel is a $7 \times 7$ matrix. Each file in the simulation dataset was generated in two versions: the first version with "no beam" (the original one, which may serve as "ground truth" in evaluating the quality of the beam reduction algorithm), and the second one, with beam effect (generated by convolving the original file with the kernel on which beam reduction is to be computed)}
\paragraph*{}{All the files produced are HI column density maps at 21cm. The average density is $1\mbox{cm}^{-3}$. The values of the compressible fraction $c_f$ vary between 0.0 (only solenoidal modes) and 1.0 (only compressible modes), with intermediate values of 0.2, 0.5, and 0.8. The amplitude of the magnetic field in each dimension $x$, $y,$ or $z$ is expressed in multiples of $7.63\mu$G, where simulations with $b_{x00}$ are hydrodynamic, and those with $b_{x05}$, $b_{x10}$, and $b_{x20}$ have magnetic field components along $x$ of $3.86, 7.63,$ and $15.26\mu$G, respectively. We also include simulation outputs at $b_{x50}$  corresponding to $bx = 38.2\mu$G, but it should be noted that this value is uncommon. Four different forcing amplitudes  were included in the initial condition parameter space of the simulation suite performed by Scholtys et al. (in prep.), namely 9000, 18000, 36000, and 72000 ($rms$), although not all combinations of $c_f$, $b_x$, and $rms$ have a corresponding simulation. Due to the complex interplay between the cooling, the turbulent velocity field, and the magnetic field, simulations with higher magnetic fields yield lower velocity dispersions than their low-$b_x$ counterparts. Therefore, simulations with $b_{x00}$, $c_f=0.5,$ and $rms$ values of 9000, 18000, and 36000 have 3D velocity dispersions of roughly 3.5, 5.2, and 8.7 km$\,$s$^{-1}$, while runs with $b_{x20}$, $c_f=0.5$ and $rms$ values of 18000, 36000, and 72000 produce 3D velocity dispersions of roughly 3.5, 5.2, and 8.7 km$\,$s$^{-1}$}.
\paragraph{}{We have at our disposal a set of stable simulation outputs:  simulations are considered stable when the global properties of the simulation have reached a steady state, that is to say that their average values no longer change as a function of time, for at least 1.5 to 2 dynamic times; the properties considered are: the velocity dispersions according to each spatial component, the mass and volume fractions occupied by each of the phases, and the distribution of temperature and magnetic field as a function of density. This indicates that gas thermodynamics and the balance between turbulence forcing and dissipation has been well established. In this study, we tested our deconvolution algorithm on data with different characteristics in forcing, initial magnetic field, and velocity distribution, and we studied the notable differences obtained on their singularity spectra.}
\paragraph{}{Three  files used in  the experiment are shown  in Table~\ref{tbl:filenames}. Each file name corresponds to three subfiles, one for each view ($x$, $y,$ and $z$).
	\begin{table}[h]
		\caption{Filenames given to the three simulation outputs.} 
		\label{tbl:filenames}
		\begin{center}
			\begin{tabular}{|c|c|c|c|} 
				\hline
				\textbf{\textit{name}}& \textbf{\textit{$c_f$}} & \textbf{\textit{$b_x$}} &  \textbf{\textit{$rms$}}                                                                                                                                     \\ 
				\hline 
				fileA                  & 0.2                 & 00 & 09000 \\ 
				\hline 
				fileB                   & 0.8                  & 00 & 09000 \\ 
				\hline 
				fileC                   & 0.5                  & 10 & 36000 \\ 
				\hline
			\end{tabular}
			\tablefoot{Each filename in the table corresponds to three subfiles, one for each projection; i.e., projection on the $y-z$ plane (called the $x$ view), projection on the $x-z$ plane (called the  $y$ view), and projection on the $x-y$ plane (called the $z$ view).}
		\end{center}
\end{table}}
\paragraph{}{As a first example, let us consider fileA in each view (i.e., $x$, $y,$ and $z$). There is no magnetic field in this case. The simulation outputs are displayed in Fig.~\ref{MHD-sim1}. We reduce the beam effect using the methodology explained in section~\ref{sec:sparse} and compute the singularity spectra of each map, that is, for the  "no beam", "beam", and with our beam-reduction algorithm (the latter is called "beam corrected" or "debeamed"). The methodology used to determine the values of the parameters $p$, $q$, and $\lambda$ is detailed in Appendix~\ref{sec:choice}; as explained there, in order to obtain a prior in the form of an $l^p$ norm, we always take $q=-1$. Values of $p$ and $\lambda$ are $p=1.7$, $p=1.5$, $p=1.4$  in $x$, $y$, and $z$, respectively, and $\lambda = 0.1$. In Figure ~\ref{spectra-MHD1} we can see the performance of the deconvolution method evaluated on singularity spectra. In the left panel of figure~\ref{sim1-spectra-x-y-z}, we show the three singularity spectra computed in the three views of data file fileA  after beam reduction. The spectra are coincident, indicating the same turbulence statistics in the three directions of space. We note that the spectra are not symmetrical, and deviate notably from a $\log$-normal behavior.}
\paragraph{}{As a second example, we consider another data file, which differs only by a modified compressive fraction $c_f$, the other parameters being the same. Figure~\ref{MHD-sim1B} shows the resulting fileB in each view. Values of $p$ and $\lambda$ are $p=1.4$, $p=1.4$, and $p=1.35$  in $x$, $y$, and $z$, respectively, and $\lambda = 0.1$. In Fig.~\ref{spectra-MHD2} we note a similar performance in  beam deconvolution. We note that the beam convolution has very little effect on the spectrum for the $x$ view. The middle panel of Fig.~\ref{sim1-spectra-x-y-z} shows the three singularity spectra computed in the three views of data file fileB  after beam reduction. There is a slight anisotropy in the $z$ view. This anisotropy, which is not observed for the more solenoidal case (fileA), might be due to a slight excess of compressive effects (compact regions), which would appear randomly and would only be visible here for one of the chosen projections.}
\paragraph{}{In a third example, we introduce an initial magnetic field and consider data fileC in the $x$, $y,$ and $z$ views (figure~\ref{MHD-sim2}). The right panel of figure~\ref{sim1-spectra-x-y-z} shows the three singularity spectra computed in the three views of the data file  after beam reduction. In this case, the anisotropy of the turbulence statistics is even more marked in the $x$ direction. We see that in the simulations, the initial magnetic field clearly results in anisotropic turbulence. This complicates observations as we only see one direction, but the result is interesting, as it shows that pure hydrodynamics simulations might not represent all the observed singularity spectra equally well and that  the magnetic field indeed influences the statistics of turbulence, as encoded in the singularity spectrum. More precisely, the curves show that, in compressive mode, the curves clearly converge towards close values when the magnetic field increases, whereas greater dispersion is observed in the more clearly solenoidal mode. Such a result is expected, and is confirmed by our analysis.}
\section{{Application to {\sl Herschel} data}}
\label{sec:hdata}
\paragraph{}{We considered observation maps from the {\sl Herschel} Gould Belt Survey~\citep{Andre2010} acquired by the SPIRE instrument~\citep{Griffin2010} at a high spatial resolution and dynamical range. We focused on the 250 $\mu$m SPIRE images to study the ISM and their embedded stellar cores~\citep{Bontemps2010,Konyves2010,Men2010}. These data files are publicly available at the Gould Belt project web page\footnote{\href{http://www.herschel.fr/cea/gouldbelt/en/Phocea/Vie_des_labos/Ast/ast_visu.php?id_ast=66}{\textcolor{blue}{Link to Gould Belt survey data page.}}}. They are very similar to the data that are also publicly available on the {\sl Herschel} Science center archive.}
\paragraph{}{In this study, we chose to examine four regions, which are described in the following subsections.  We retake the dense but currently nonstar-forming filament Musca from our previous study. In addition, we aim to extend our study to other interstellar clouds: specifically, very diffuse clouds (Spider)\footnote{Spider is not in the {\sl Herschel} Gould Belt Survey database.} without dense filaments, a nearby low-mass star-forming region (Taurus), and a nearby star-forming region hosting the impact of stellar feedback (Aquila). We also mention our goal to study a larger sample of star-forming regions in future work and to extend to high-mass star-forming regions. For this purpose, we seek to demonstrate the general character of the image processing methods introduced here and in our previous article, by showing that they apply without difficulty to a large number of types of observation maps. Not yet focusing on high-mass star-forming regions also makes sense because they are a very specific topic in the current debate of ISM evolution. In the numerical implementation of the model presented in section~\ref{sec:sparse}, the beam kernel of the SPIRE instrument is a $20 \times 20$ matrix obtained by calibration on Neptune~\citep{Griffinetal2010,Vatchanov2017}.}
\subsection{Effect of beam reduction on the Musca observation map}
\begin{figure}[h]
\centering
\includegraphics[scale=0.5]{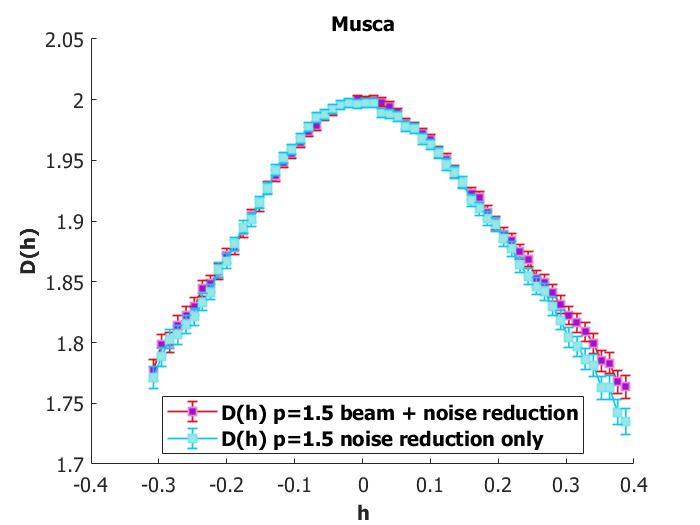} 
\caption{Beam effect on Musca. On the Musca map, the beam effect does not influence the computation of the singularity spectrum.}
\label{musca-debeam}
\end{figure}
\paragraph{}{On the Musca observation map, the effect of beam deconvolution is negligible compared to noise reduction. Indeed, the model for noise and beam reduction presented in section~\ref{sec:sparse} can be implemented using the beam kernel matrix ${\bf H}$, which is defined in the case of the SPIRE instrument by the previously mentioned matrix obtained by calibration on Neptune. In this case, beam deconvolution and noise reduction are both operated, but we can also  set ${\bf H} = {\bf Id}$ (identity matrix), in which case only noise reduction is done. Figure~\ref{musca-debeam} shows the resulting singularity spectrum in the case where $p = 1.5$, $q = -1$, and $\lambda = 0.1,$ with and without beam deconvolution. The curves are very similar, indicating that, at least in the case of Musca, the beam effect is not a hurdle in the multifractal analysis of that map.}
\subsection{Aquila rift}
\begin{figure}[h]
\centering
\includegraphics[width=0.5\textwidth]{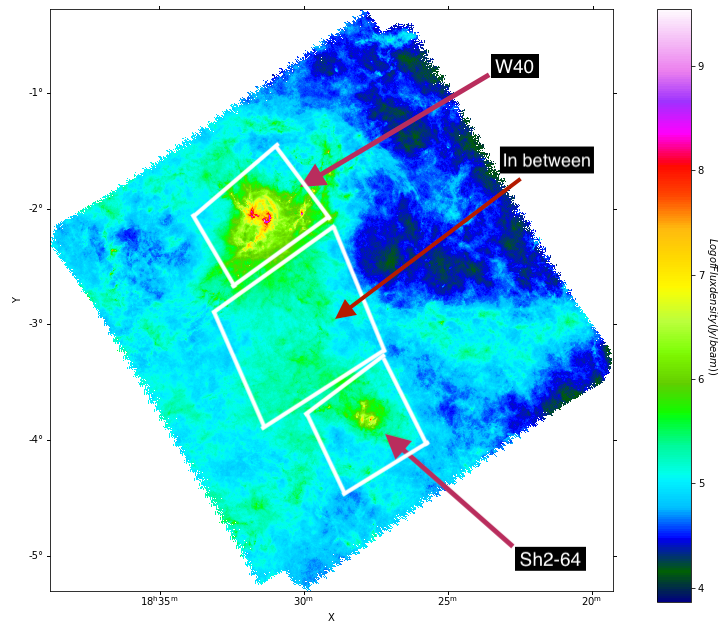} 
\caption{Aquila flux intensity map from {\sl Herschel} at 250 $\mu$m. }
\label{aquila-show1}
\end{figure}
\begin{figure}[h]
\centering
\includegraphics[width=0.5\textwidth]{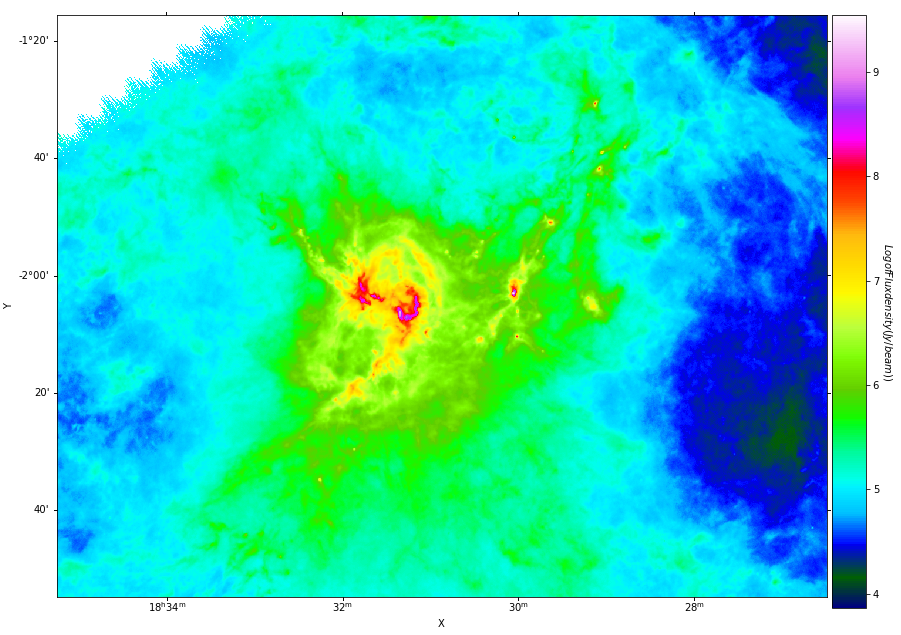} 
\caption{Aquila flux intensity map from {\sl Herschel} at 250 $\mu$m.  We show a zoom onto one subregion where the flux intensity is more intense (W40 subregion).}
\label{aquila-show2}
\end{figure}
\begin{figure*}[h]
\centering
\includegraphics[width=0.3\textwidth]{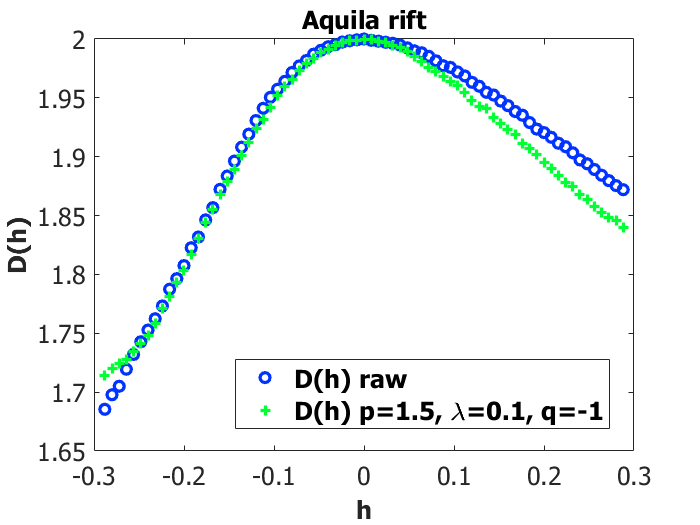}  \includegraphics[width=0.3\textwidth]{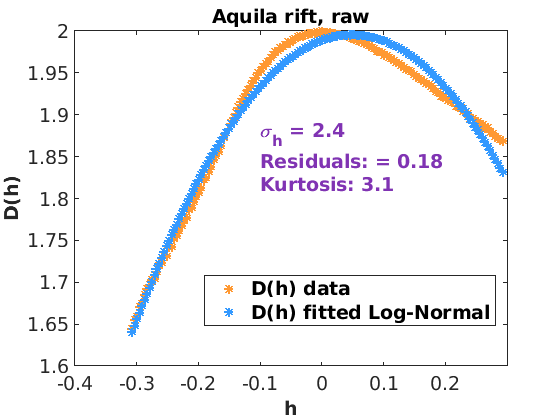}  \includegraphics[width=0.3\textwidth]{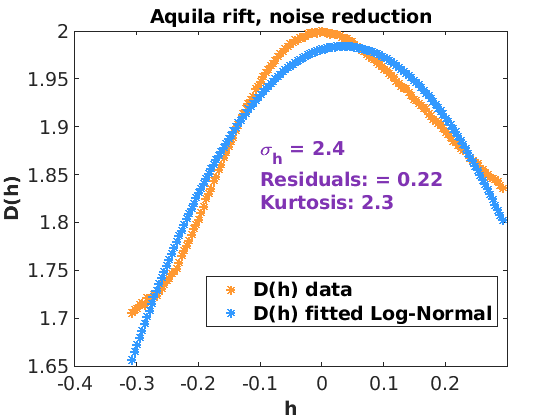} 
\caption{Aquila: singularity spectrum  and noise reduction. Left: Singularity spectra of Aquila observation map with and without noise reduction. Middle: $\log$-normal fit with raw data. Right: $\log$-normal fit after noise reduction. The filtered observation map is less Gaussian than the original raw data.}
\label{aquila-spectra}
\end{figure*}
\begin{figure*}[h]
\centering
\includegraphics[width=0.3\textwidth]{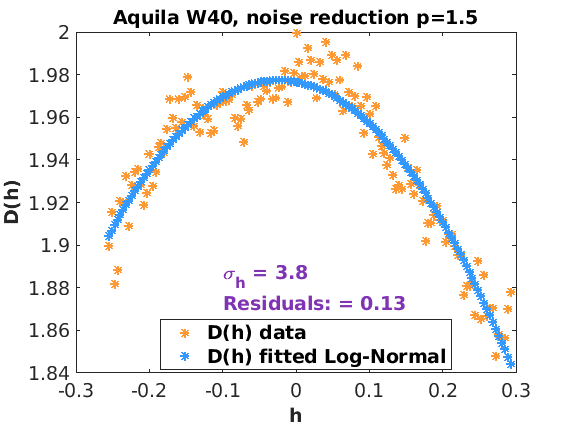}  \includegraphics[width=0.3\textwidth]{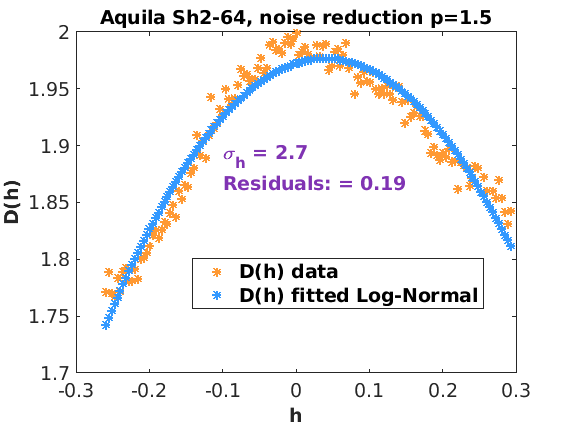}  \includegraphics[width=0.3\textwidth]{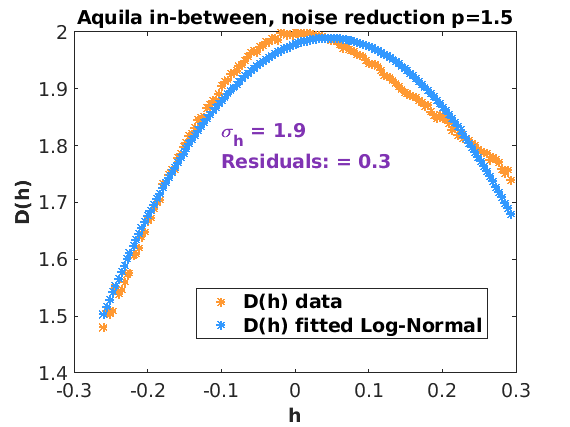} 
\caption{In Aquila, subregions W40 and Sh2-64 are more $\log$-normal than the region in between.}
\label{ln-aquila-regions}
\end{figure*}
\begin{figure}[h]
\centering
\includegraphics[width=0.5\textwidth]{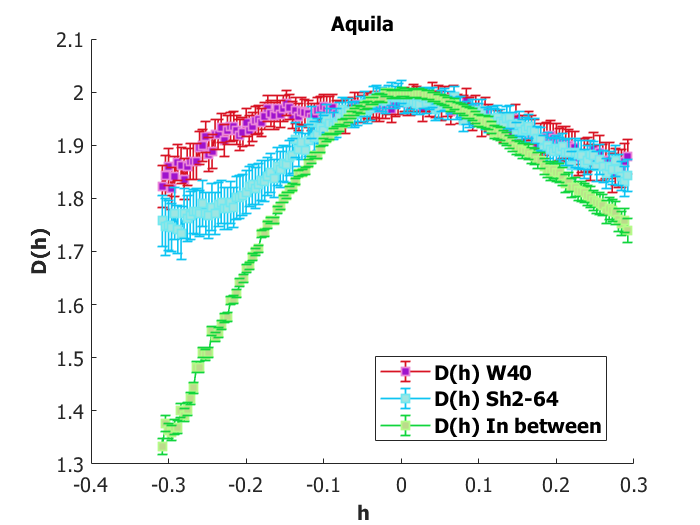} 
\caption{Singularity spectra computed over W40, Sh2-64 and in the area between the two.}
\label{spectra-aquila-regions}
\end{figure}
\begin{figure*}[h]
\centering
\includegraphics[width=0.44\textwidth]{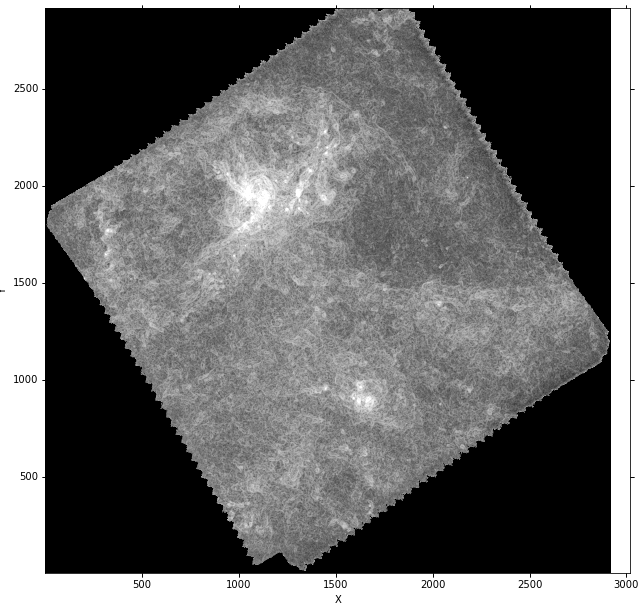}  \includegraphics[width=0.55\textwidth]{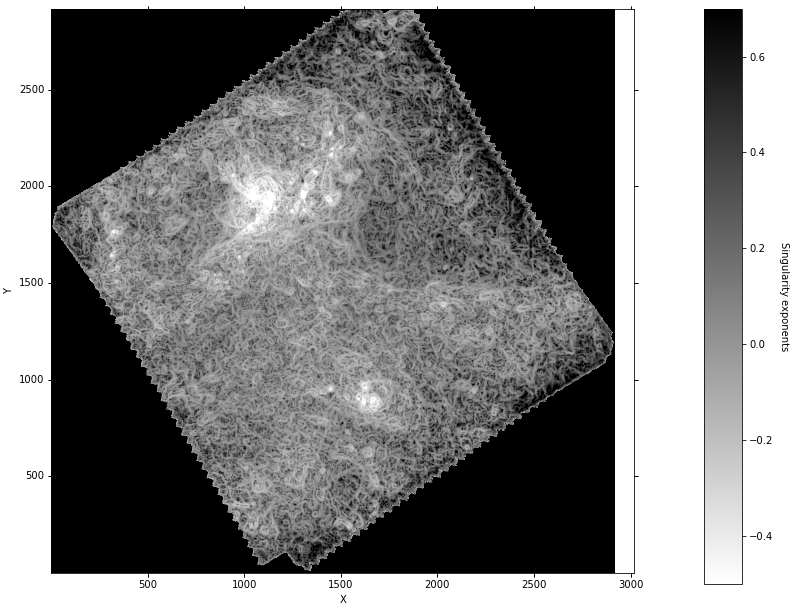} \\
\includegraphics[width=0.85\textwidth]{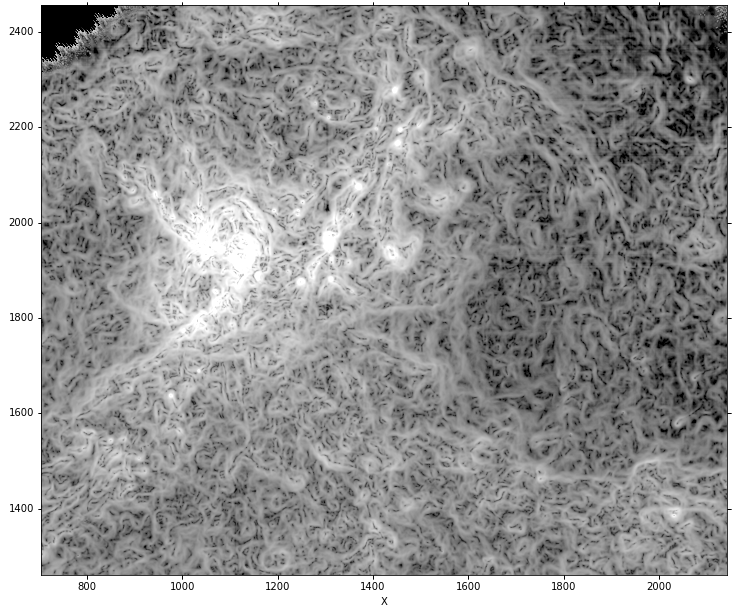} 
\caption{Noise reduction on Aquila. Top panel: singularity exponents of the {\sl Herschel} observation of Aquila at 250 $\mu$m. Left image:  display of the singularity exponents of the raw, unfiltered map. Right image: after noise reduction, $p=1.5$, $q=-1$, $\lambda=0.1$. Bottom panel: zoom on subregion of the filtered image.}
\label{aquila-se}
\end{figure*}
\paragraph{}{The Aquila rift is a structure of  $5^{\circ}$   in length situated above the Galactic plane at $l=28^{\circ}$ and located at an uncertain distance ranging from $240$ to $500$ pc~\citep{Comeron2022}. As pointed out in~\citet{Bontemps2010}, {\sl Spitzer} observations indicate the existence of an embedded cluster in the Aquila rift referred to as the Serpens South cluster,  which belongs to the Serpens molecular cloud. Figure~\ref{aquila-show1} displays the $\log$ of the flux intensity of the Aquila rift at 250 $\mu$m, with a focus on the HII region shown in figure~\ref{aquila-show2}. Figure~\ref{aquila-se} (top panel) shows the singularity exponents ---corresponding to the local correlation measure of formula (B.7) in~\citep{yahia2021}--- of the raw  Aquila 250 $\mu$m (top panel), and the map after noise reduction with $p=1.5$, $q=-1$, and $\lambda=0.1$ (bottom panel). We note that there is more similarity between the two maps than we observed with the Musca map. This comes from the fact that we see less of the background noise for this region as the average column density of dust is larger towards Aquila than towards Musca and then dominates over the background. In the bottom panel of figure~\ref{aquila-se}  we zoom onto the central part of the filtered Aquila map. The image of the singularity exponents reveals a similar complex web of smaller-scale filamentary structures, which extend the whole area in a spectacular way. In the left panel of figure~\ref{aquila-spectra}, we see that the singularity spectra of the raw map and the filtered map agree on the most singular transitions of the signal (i.e., ${\bf h} \leq 0$), which confirms that there is less background noise. However, we observe that the two spectra depart from each other in the range ${\bf h} > 0$. The two other image panels show the result of a $\log$-normal fit of the singularity spectra $D({\bf h})$. The middle panel of figure~\ref{aquila-spectra} displays the result of the $\log$-normal fit for the raw Aquila data, and the right image shows the  $\log$-normal fit for the filtered Aquila data with parameters $p=1.5$, $q=-1$, and $\lambda=0.1$.  }
\paragraph{}{Figure~\ref{spectra-aquila-regions} displays the singularity spectra computed over W40, Sh2-64, and over the region between the two. Star-forming regions W40 and Sh2-64 depart clearly from the in-between region. This is confirmed by the $\log$-normal fits of each region, with the star-forming regions being more $\log$-normal as shown in figure~\ref{ln-aquila-regions}. The range of ${\bf h}$ values used in generating figure~\ref{ln-aquila-regions} has been restricted to $[-0.25, 0.3]$ because of the reduced size of the  W40 and  Sh2-64 subregions, and the comparison to a $\log$-normal process is based on the residuals for the same reason. This behavior corresponds very well to what was observed in Musca with a clear tendency to show a more $\log$-normal behavior in the densest regions (especially in the upper part of Musca, where we observe the formation of a protostar, as well as toward the crest of the filament) as discussed in~\citep{yahia2021}. We thus observe a phenomenon similar to that observed on Musca: the regions containing the densest filaments have a more $\log$-normal singularity spectrum than the less dense regions. The singularity spectrum of the region located between W40 and Sh2-64 (in green in Figure~\ref{spectra-aquila-regions}) clearly shows a preponderance of filamentary regions of lower dimensions, a phenomenon also observed on Musca in the less dense regions.}
\subsection{Taurus L1495}
\begin{figure*}[h]
\centering
\includegraphics[width=0.8\textwidth]{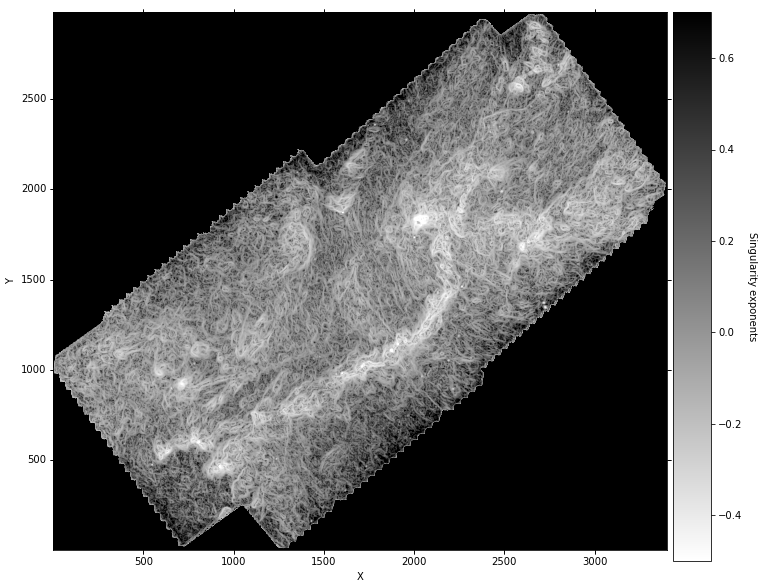} 
	\includegraphics[width=0.9\textwidth]{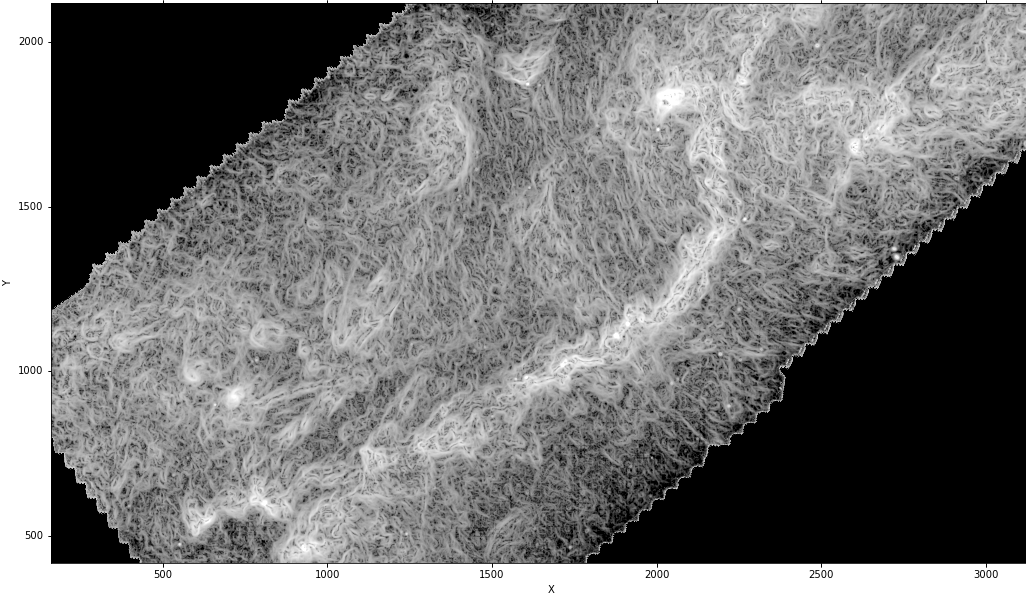}
\caption{Noise reduction on Taurus L1495. Top panel: Singularity exponents of the {\sl Herschel} observation of Taurus L1495 at 250 $\mu$m after noise reduction: $p=1.5$, $q=-1$, $\lambda=0.1$. Bottom panel: Zoom onto southern subregion.}
\label{taurus-show}
\end{figure*}
\begin{figure*}[h]
\centering
\includegraphics[width=0.3\textwidth]{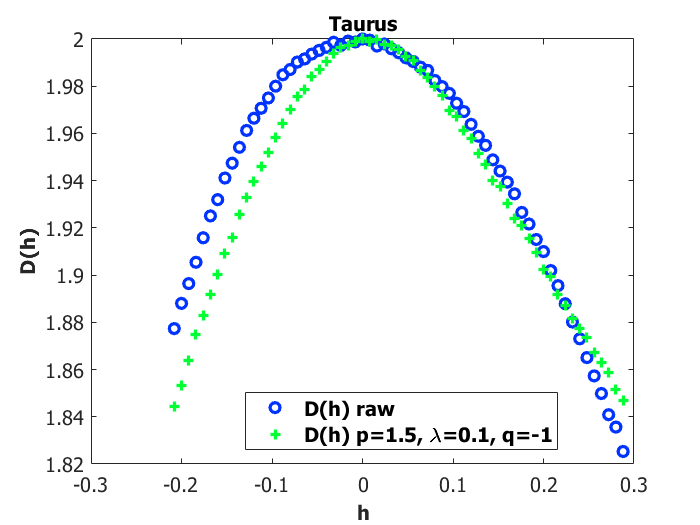} \includegraphics[width=0.3\textwidth]{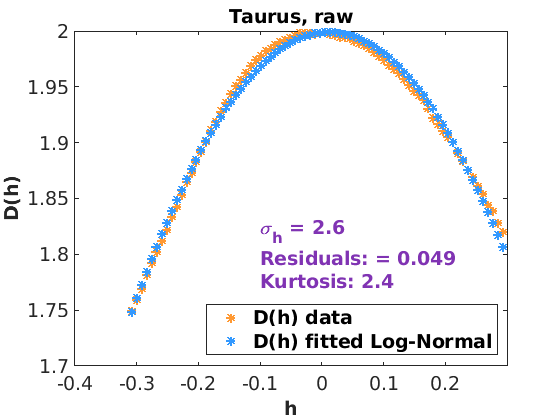}  \includegraphics[width=0.3\textwidth]{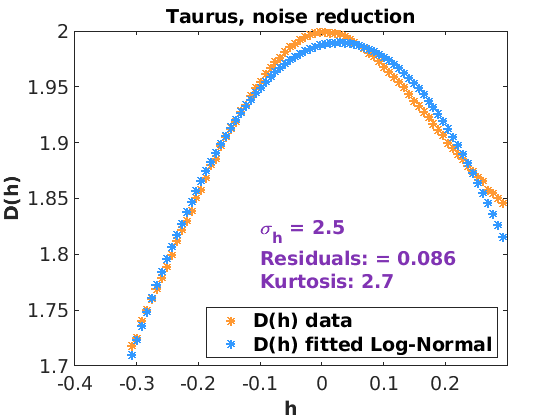} 
\caption{Taurus: singularity spectrum and noise reduction. Left panel: Singularity spectra of Taurus L1495 observation map with and without noise reduction. Middle panel: $\log$-normal fit to raw data. Right panel: $\log$-normal fit after noise reduction: $p=1.5$, $q=-1$, $\lambda=0.1$.}
\label{taurus-spectra}
\end{figure*}
\begin{figure}[h]
\centering
\includegraphics[width=0.5\textwidth]{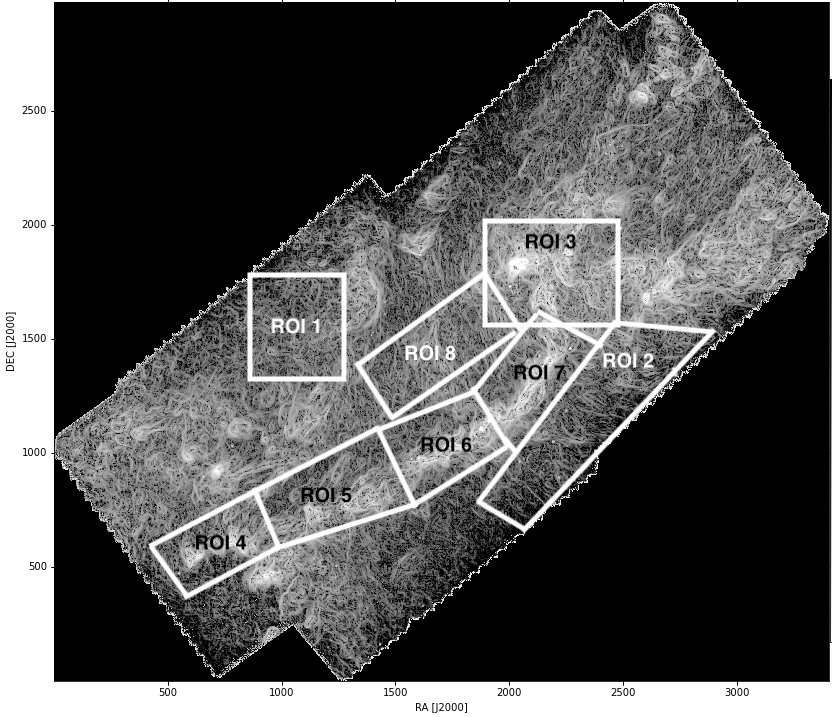}
\caption{Regions of interest defined over the Taurus observation map.}
\label{taurus-rois}
\end{figure}
\begin{figure*}[h]
\centering
\includegraphics[width=0.236\textwidth]{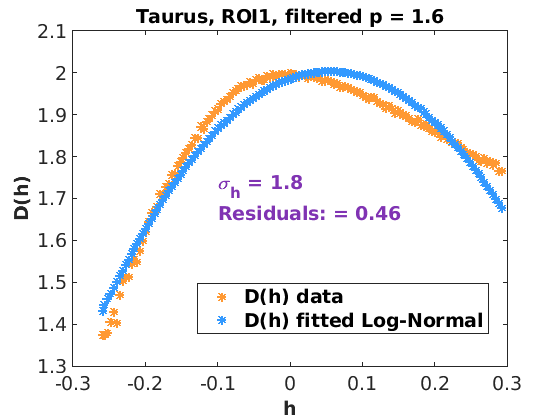} \includegraphics[width=0.236\textwidth]{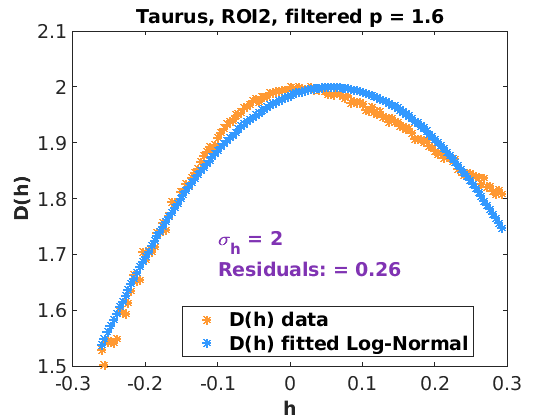} \includegraphics[width=0.236\textwidth]{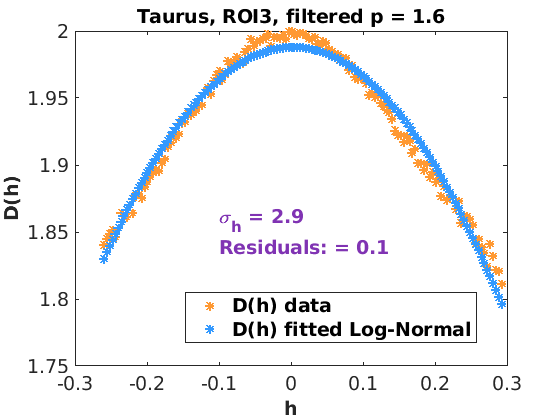} \includegraphics[width=0.236\textwidth]{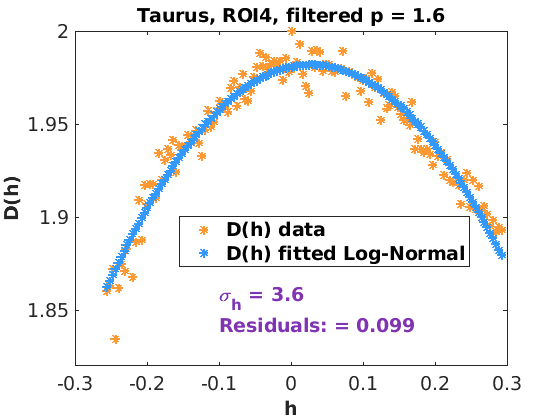}  
\includegraphics[width=0.236\textwidth]{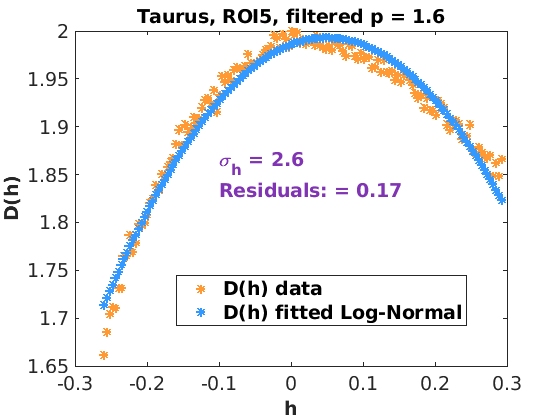} \includegraphics[width=0.236\textwidth]{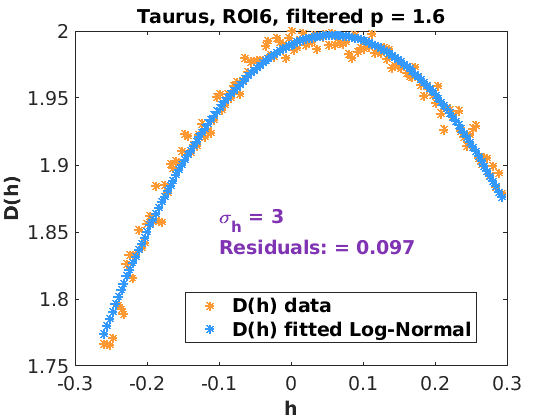} \includegraphics[width=0.235\textwidth]{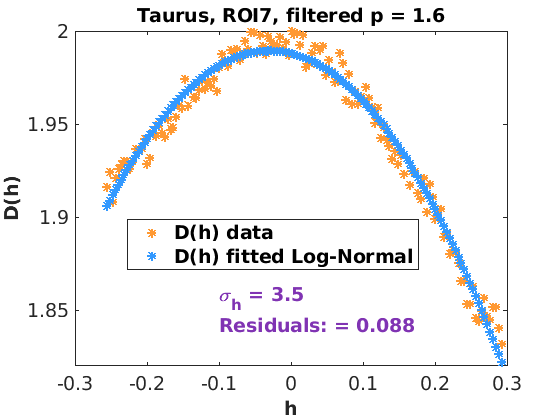} \includegraphics[width=0.235\textwidth]{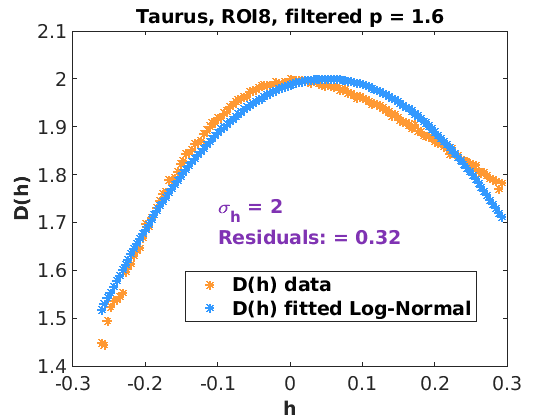}
\caption{Fit of the singularity spectra for each ROI (observation map filtered with $p=1.6$) by a $\log$-normal spectrum. Top row: ROIs 1 to 4, bottom: ROIs 5 to 8.}
\label{taurus-rois-lognf}
\end{figure*}
\begin{figure*}[h]
\centering
\includegraphics[width=0.45\textwidth]{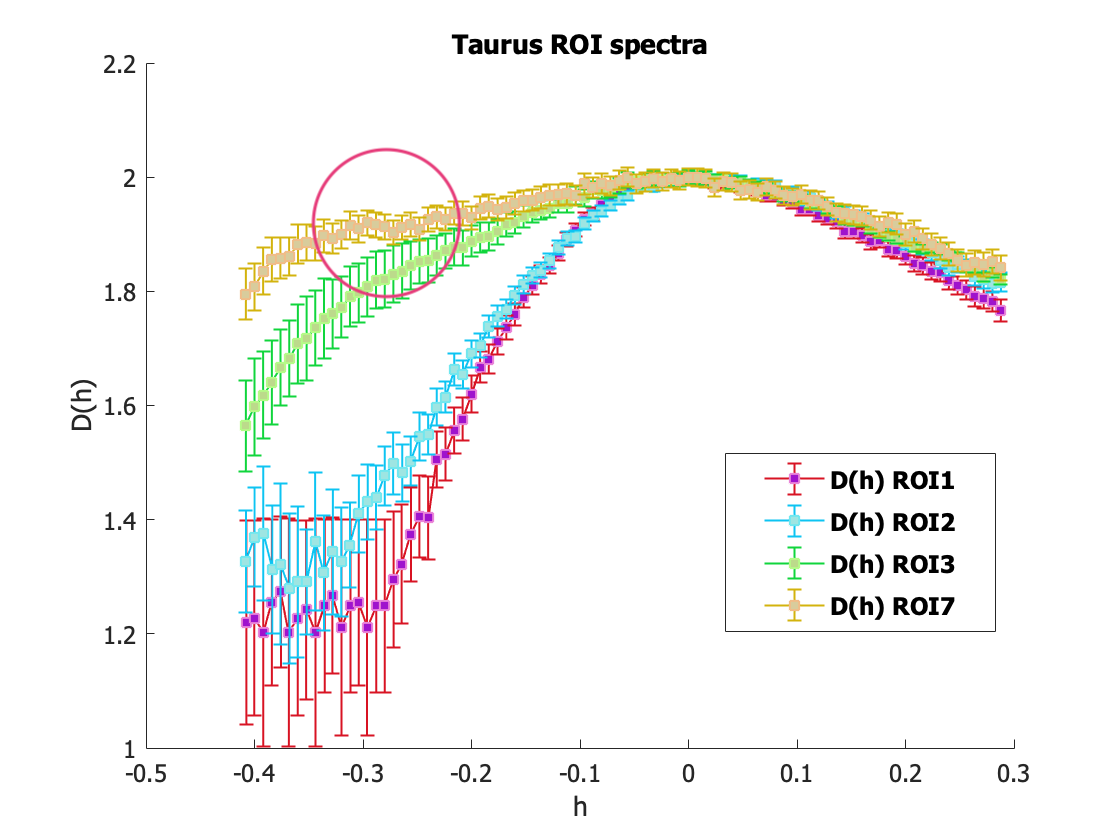}  \includegraphics[width=0.45\textwidth]{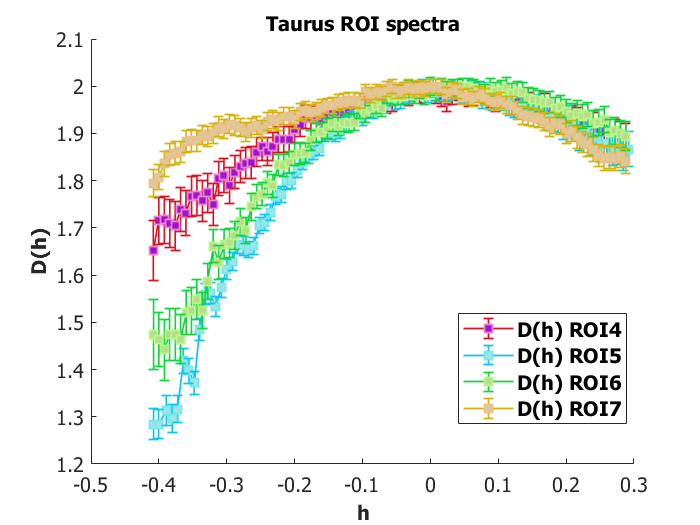} 
\caption{Different turbulent dynamics in the ROIs. Left panel: Regions 1, 2, 3, and 7. Right panel: Regions 4, 5, 6, and 7. The red circle in the left panel is placed over a possible inflexion point in the spectrum, which indicates at least two different types of turbulent dynamics in ROI 7, a finding that can be compared to the results of~\citet{Li2021}}
\label{taurus-spectrum-eb-ROI7}
\end{figure*}
\begin{figure}[h]
\centering
\includegraphics[width=0.5\textwidth]{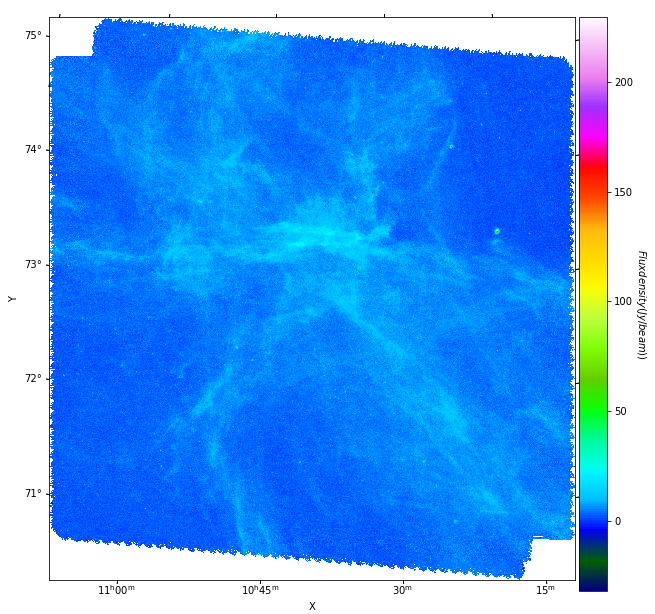}
\caption{Spider flux intensity map from {\sl Herschel} at 250 $\mu$m.}
\label{spider-show}
\end{figure}
\begin{figure}[h]
\centering
\includegraphics[width=0.4\textwidth]{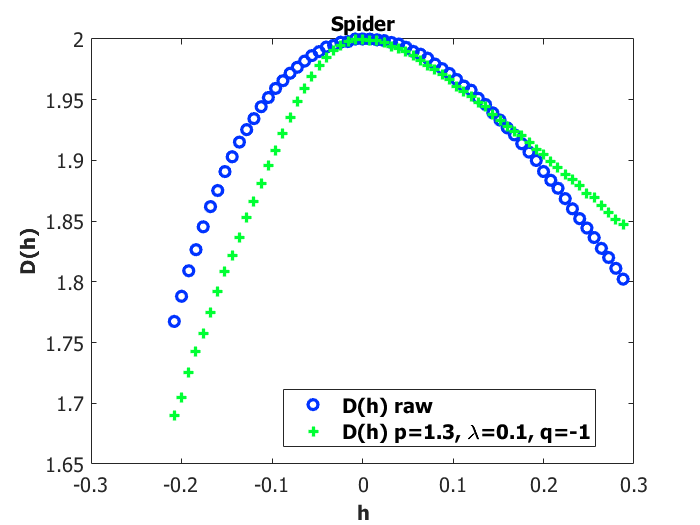} 
\includegraphics[width=0.24\textwidth]{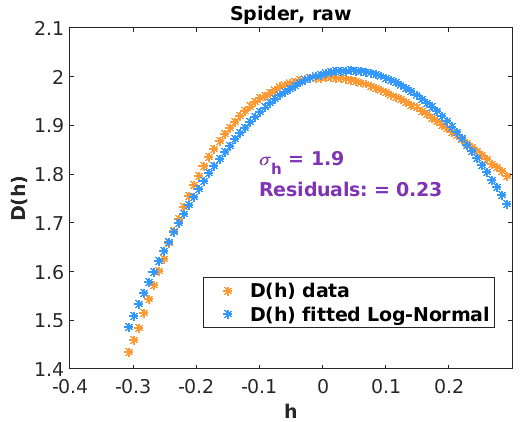}  \includegraphics[width=0.24\textwidth]{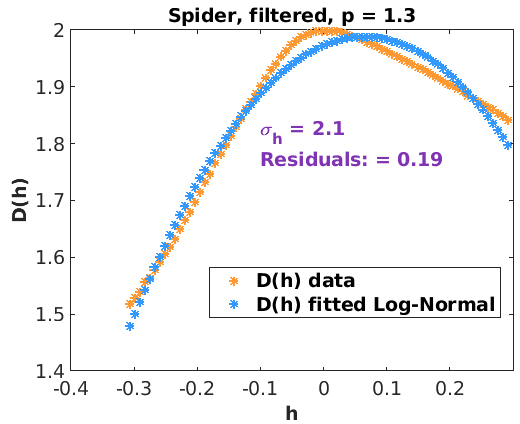} 
\caption{Noise reduction on Spider. Top: Singularity spectra of Spider observation map with and without noise reduction. Botton left: $\log$-normal fit to raw data. Botton right: $\log$-normal fit after noise reduction: $p=1.3$, $q=-1$, $\lambda=0.1$.}
\label{spider-spectra}
\end{figure} 
\begin{figure*}[h]
\centering
\includegraphics[width=0.45\textwidth]{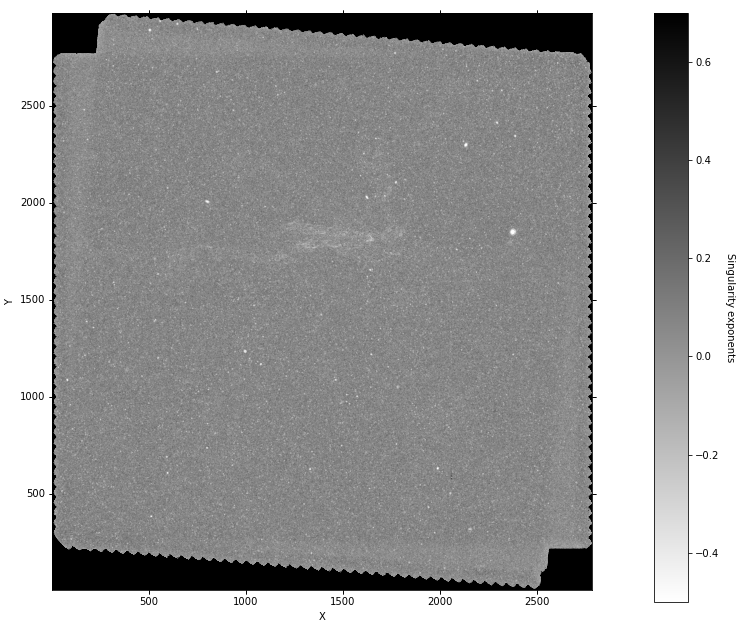}  \includegraphics[width=0.45\textwidth]{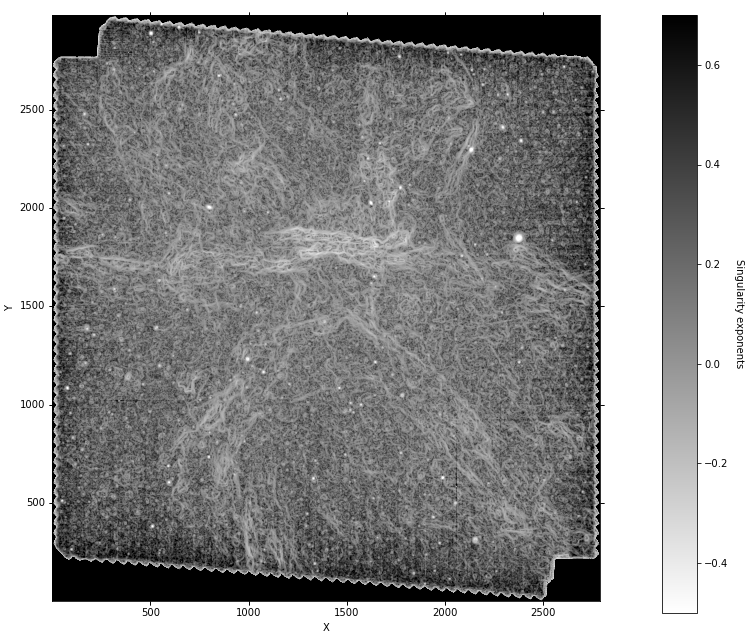} 
\includegraphics[width=0.9\textwidth]{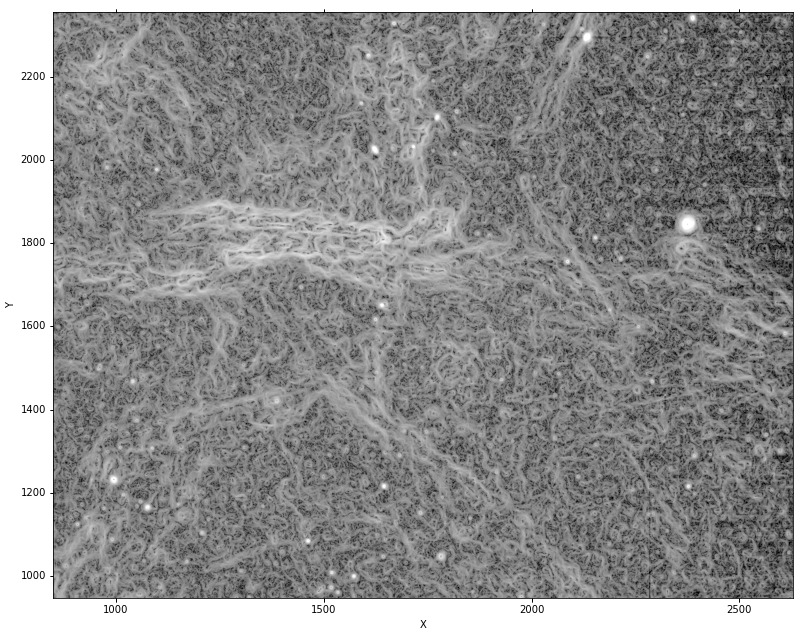}  
\caption{ Spider: singularity spectrum and noise reduction. Top panel: Singularity exponents of the {\sl Herschel} observation of Spider at 250 $\mu$m. Left:  Display of the singularity exponents of the raw, unfiltered map. Right: After noise reduction: $p=1.3$, $q=-1$, $\lambda=0.1$. Bottom panel: Zoom onto subregion (filtered map).}
\label{spider-se}
\end{figure*}
\begin{figure*}[t]
\centering
\includegraphics[width=0.3\textwidth,height=0.35\textheight]{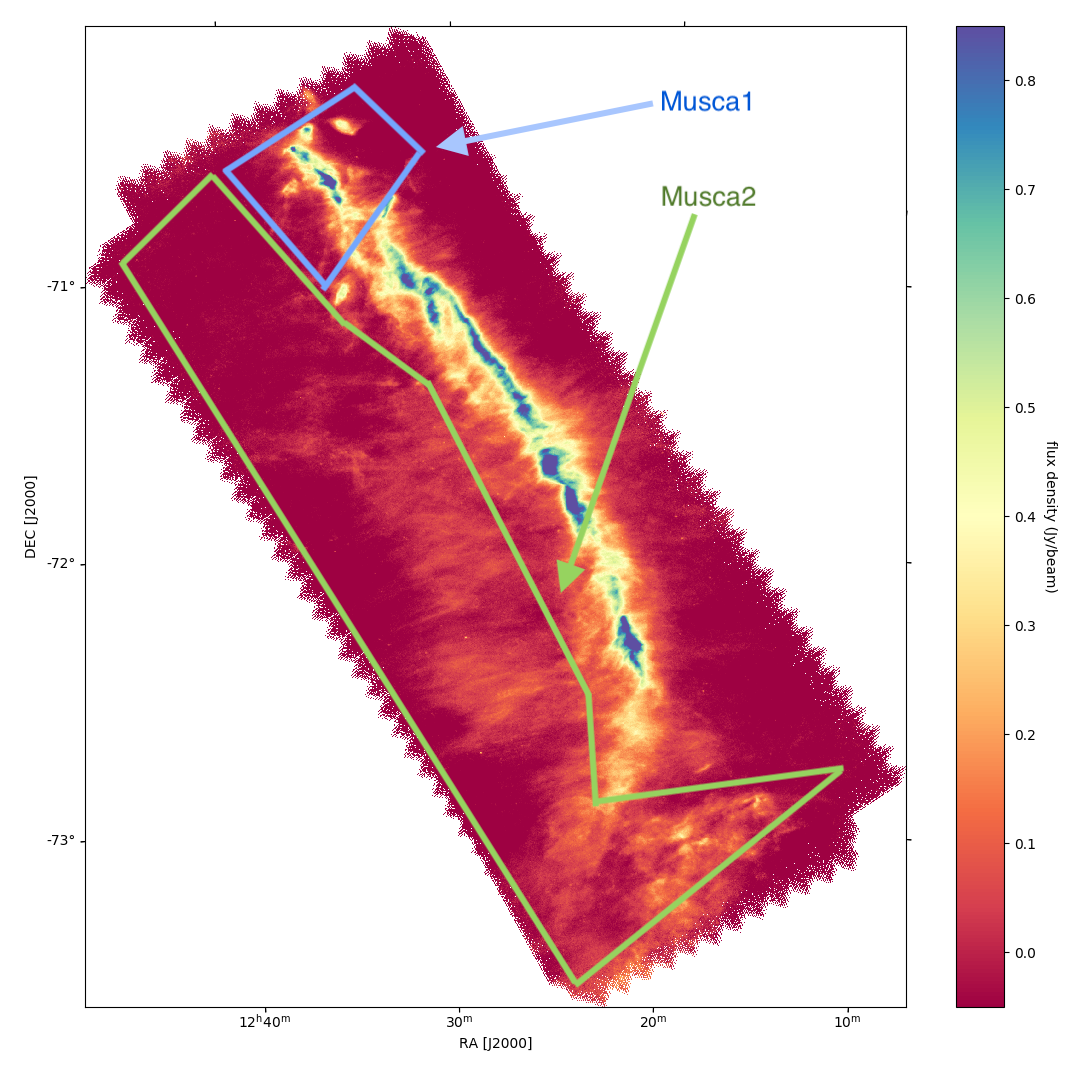}  \includegraphics[width=0.67\textwidth]{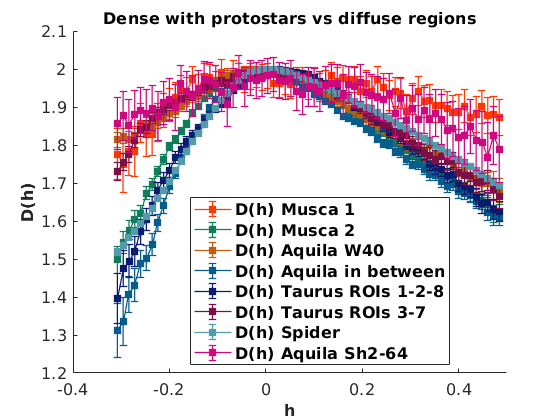}  
\caption{Dense regions with protostars vs diffuse regions. Left panel: Definition of subregions Musca 1 and Musca 2. Right panel: Left parts of singularity spectra (i.e., corresponding to ${\bf h} \leq 0$) clearly show two classes associated with diffuse regions and dense regions containing protostars.}
\label{spectra-classification}
\end{figure*}
\paragraph{}{The Taurus star-forming region has been documented in a number of studies using the {\sl Herschel } SPIRE and PACS data~\citep[e.g.,][]{Kirk2013,Palmeirim2013,Marsh2016,Roy2019}. It has been noted that the star-forming filament B211, in a way very similar to Musca, reveals the presence of striations perpendicular to the filament and that these striations are generally oriented along the magnetic field direction~\citep{Palmeirim2013,PlanckArzoumanian2016}. Based on complementary kinematic observations,~\citet{Palmeirim2013} and~ \citet{Shimajiri2019} proposed that this filament is gravitationally accreting background cloud material from the molecular cloud, whereas \citet{Bonne2020b} proposed the mass inflow might be mostly driven by magnetic field bending, which is similar to what was proposed for Musca.}
\paragraph{}{The top panel of Fig.~\ref{taurus-show}  displays the image of the singularity exponents covering the whole SPIRE 250$\mu$m observation map filtered with values $p=1.5$, $q=-1$, and $\lambda=0.1$, and the bottom panel shows a zoom onto the southern subregion. The complexity of small-scale filamentary structures is clearly visible. Figure~\ref{taurus-spectra} demonstrates, as in the Musca observation map, the positive effect of noise reduction (here seen quantitatively in the value of the residuals): elimination of the Gaussian noise, notably on the most negative singularity exponents (corresponding to small-scale filamentary structures), clears spurious $\log$-normality in the spectrum not related to the dynamics of the medium, but coming from noise. }
\paragraph{}{In Fig.~\ref{taurus-rois} we define eight regions of interest (ROIs) in the {\sl Herschel} SPIRE 250$\mu$m, some of them related to the regions highlighted in~\citet{Hacar2013} and \citet{Li2021}. The regions are chosen at various locations inside the main filament and outside of it, in some less dense regions where smaller-scale striations prevail. We select these regions because ROIs 3, 4, and 6 host protostars, ROIs 5 and 7 do not host protostars (yet), and ROIs 1, 2, and 8 trace regions outside the dense gas.  We now compare our results obtained with the computation of the singularity spectra on the noise-reduced observation maps with the results  of~\citet{Hacar2013} and \citet{Li2021}.  Figure~\ref{taurus-rois-lognf} displays, for each region, the result of a $\log$-normal fit of each singularity spectrum. Considering the residual $\varepsilon_i$ of ROI $i$, we see that ROI 7 is the closest to $\log$-normality, followed by ROI 6, ROI 4, ROI 3,  and ROI 5. On the other hand, regions 1, 2, and 8 (which are outside of the areas studied by \citet{Hacar2013}) depart clearly from the previous group; they are made up of filaments that seem to blend together less, which  results in a greater number of isolated filaments contributing much less to the formation of stable zones. Indeed Kolmogorov's $\log$-normal model is associated with the presence of large chunks of nearly equal dissipation rates. The values of the error fit for regions 1, 2, and 8 indicate a stronger deviation from log-normality than other ROIs. Thus, the phenomenon observed for the ROIs 1, 2, and 8 is similar to that observed for the intermediate region between Sh2-64 and W40 in Aquila. (We reiterate that a residual is a root mean square error (RMSE) between singularity spectra computed on the data and a $\log$-normal fit: smaller residuals imply a more $\log$-normal behavior). We note that, in figure~\ref{taurus-rois-lognf}, the small size in pixels of these regions imposes that we restrict the interval of ${\bf h}$ values to $[-0.25, 0.3]$ and we do not make use of the kurtosis because high statistical moments are difficult to interpret when there are few data (e.g., the common boundary between ROI4 and ROI5 is about 200 pixels).  Figure~\ref{taurus-spectrum-eb-ROI7} shows a comparison between the spectra of different regions. The left panel displays the singularity spectra of regions  1, 2, 3, and 7. We clearly see that regions 1 and 2 have similar spectra, while the spectra of regions 3 and 7 depart from the other two. The right panel shows the spectra of the regions 4, 5, 6, and 7, which lie on the main filament of Taurus. While regions 5 and 6 have close spectra, we note marked differences between these regions. The red circle on the left panel of the figure indicates a possible inflexion point in the  spectrum of ROI 7. Although this is probably a second-order effect, it is questionable whether this observation can be compared with the findings of~\citet{Li2021} of the existence of two regions driven by different dynamics: one subregion dominated by gravity and one by the magnetic field; however,  as region ROI 7 is relatively small in pixel size, further confirmation is required in this regard. If confirmed, this would underline the usefulness of the singularity spectrum as a tool to detect underlying dynamics.}
\paragraph{}{We also observe that ROIs 3, 4, 5, 6, and 7 (in the dense gas) are better fitted by $\log$-normal spectra than ROIs 1, 2, and 8. It thus appears that $\log$-normal behavior increases when probing the dense gas (n$_{H_{2}} >$ 10$^{4}$ cm$^{-3}$). This is in line with the conclusion we draw in the previous section for Aquila and with Musca results too. Clearly, the dense regions have similar spectra, which are more $\log$-normal and less filamentary than those of the more diffuse gas regions (outside the dense structures). This may appear slightly contradictory, since the well-observed dense structure appears as a clear large-scale filament (such as in Musca). This suggests that, despite being a large-scale filament, the dense structure is ---statistically on all scales--- less filamentary than the less dense gas surrounding it and in particular the striation regions.  The very low-intensity striations are probably the reason for these $\log$-normal deviations and for the higher filamentary level in the spectra.}
\subsection{Spider}
\paragraph{}{The Spider ISM was studied by the Planck Collaboration~\citep{PlanckCollab2011}. \citet{Barriault2010} studied the transition from atomic to molecular gas in the Spider region, and indicate that the Spider region has very low column densities of only $\sim$4 10$^{20}$ cm$^{-2}$ and shows no star-formation activity; and no molecular core was detected. Figure~\ref{spider-show} shows the {\sl Herschel} flux intensity map, and the top panel of figure~\ref{spider-se} shows the singularity exponents on the raw observation map (left) and after noise reduction using  $p=1.3$, $q=-1$, and $\lambda=0.1$. We note that proper filtering is required in such noise-dominated regions. The bottom panel of figure~\ref{spider-se} is a zoom onto the central subregion of the filtered map. Again we note the ubiquity of filamentary structures at different scales. Figure~\ref{spider-spectra} displays the singularity spectra obtained with and without noise reduction, and the result of a $\log$-normal fit. We conclude that the Spider region is highly dominated by thinner filamentary structures, which cause its spectrum to strongly deviate from a $\log$-normal behavior; this is again very much in line with previous results for Aquila, Musca, and Taurus, with a typical spectrum of diffuse gas, such as in ROI 1, 2, and 8 in Taurus for instance.}
\subsection{Dense regions with protostars versus less dense filamentary regions}
\label{sec:mainresult}

\paragraph*{}{Figure~\ref{spectra-classification}    
shows one of the most important results of this study, and one that demonstrates the strength of this approach for studying the ISM: by drawing the singularity spectra of the various regions studied  on the same graph, and by retaining dense regions containing protostars as well as diffuse regions rich in filaments of smaller scale, we clearly see the appearance of two classes  defined by the most singular part of the spectra (i.e., defined by exponent values ${\bf h} \leq 0)$: these correspond to the dense regions with protostars versus diffuse regions.}
\section{Conclusion and summary}
\label{sec:discussion}
\paragraph*{}{In this study, we present significant improvements in our approach to tackling the problem of noise reduction in the {\sl Herschel} data. We carried out a sparse filtering of the type $l^2$-$l^p$, which makes it possible to reduce the Gaussian noise while preserving the filamentary structures that appear to be universally observed in the turbulence of interstellar clouds. This article follows on from our previous study, which focuses on Musca only, where we introduced $l^1$-$l^1$ filtering. The new algorithm presented here allows noise reduction and deconvolution (to reduce the beam effect) in a single pass. The algorithm depends on several parameters, which control its convergence as well as the quality of the result. We propose a set of key values for these parameters ---which are suitable for all the {\sl Herschel} data that we have analyzed--- and a trade-off between the sparsity term and the data fit in an optimization problem. This type of algorithm was very popular before the arrival of methods based on deep-learning. The latter would undoubtedly make it possible to obtain results of comparable quality (according for instance to Peak Signal-to-Noise Ratio (PSNR)), but it is preferable in our opinion, and in an astrophysical framework, to rely on configurable algorithms, even if, as in the present case, the choice of parameters in a nonlinear context, with few data and without a ground truth,} is always difficult.
\paragraph*{}{Among the results obtained, we first note that the beam effect seems to have a negligible influence on the calculation of singularity spectra, which is unlike the problem of noise reduction, a step absolutely necessary for obtaining singularity spectra that can be used in the context of determination of distinct turbulent phenomena.}
\paragraph*{}{The less dense regions observed in the studied ISM clouds are found with less $\log$-normal and more filamentary singularity spectra than the densest regions. This may be put in pers\-pective with the proposed scenario of dynamical interactions between two HI and low-density CO clouds at significantly different velocities as proposed for Musca in~\cite{Bonne2020b}, and in DR21/Cygnus X in~\cite{Schneider2023} and~\cite{Bonne2023}.}
\paragraph*{}{We therefore conclude that the singularity spectra computed in a microcanonical formulation can be used to study  the fila\-mentary structure of the ISM  with similar spectra for low-column-density regions almost devoid of CO gas and for dense structures (mostly CO) in the regions  targeted so far (Musca, Aquila, Taurus, and Spider). These similar spectra indicate that low-intensity  regions are much less $\log$-normal and more filamentary than the dense regions.}
\paragraph*{}{We see an increased $\log$-normality of the spectra in regions with higher intensity. The new results toward Aquila, Taurus, and Spider confirm our proposals of our previous article: we find many more filamentary spectra outside the dense regions (even the dense regions that are nevertheless filamentary to the eye). This suggests that the low-intensity ISM (the HI?) is very strongly filamentary; in any case, even more so than the dense filaments that we see clearly by visual inspection.}
\paragraph*{}{Then, in addition to the clear demonstration of filamentary structures at different scales in all our observation
data, including in the least dense clouds, we were able to extend and generalize ---thanks to an efficient Gaussian noise reduction--- the possibility to calculate singularity spectra in distinct subregions. In the case of Taurus for example, we raise the question of obtained signatures of distinct turbulent processes.  The positive results obtained here demonstrate the strength of our approach, and give us confidence in the path we have chosen to elucidate the different properties of the magnetohydrodynamic turbulence in interstellar clouds, and their relationships with the complex processes of star formation.  The results presented here with real data reveal visible differences in the singularity spectra between dense areas containing cores, and less dense, more filamentary regions, which we interpret in terms of deviation from $\log$-normality. Our future work concerning MHD simulations will focus on the influences of the magnetic field and gravity on the shape of the spectra.}
\pagebreak
\begin{acknowledgements}
This work is supported by:
\\ 1) the {\it GENESIS} project (\emph{GENeration and Evolution of Structure in the ISm}), via the french ANR and the german DFG through grant numbers ANR-16-CE92-0035-01 and DFG1591/2-1. N.S. and R.S. acknowledge support by the Collaborative Research Center 1601 sub-project B2, funded by the DFG, German Research Foundation – 500700252 \\\\
2) the INRIA InnovationLab {\bf Geostat}-{\bf I2S} (Innovative imaging solutions). A. Rashidi was funded by a CIFRE PhD grant in the framework of the INRIA InnovationLab {\bf Geostat}-{\bf I2S}. 
\\\\
The authors thank the anonymous referees for indications which improved the results.
\end{acknowledgements}
\pagebreak
\bibliographystyle{aa} 
\bibliography{debeam-deconvolution-denoise-fv.bib} 
\begin{appendix}
\section{Solution method}
\label{sec:sol}
\paragraph{}{The optimisation problem of eq.~\ref{eqn:math}, that is,
\begin{equation*}
\hat{\vs} =  \argmin{\vs} \vf(\vs,\vy),~\mbox{with}~~\vf(\vs,\vy) = \| {\bf H} \vs-\vy \|^2+\lambda \phi({\cal D }\vs),
\end{equation*}
is of the form 
\begin{equation}
\hat{\vs}=  \displaystyle \argmin{\vs} \left( \vl(\vs)+\vm(\vs) \right),
\label{eqn:argmin}
\end{equation}
with
\begin{equation}
\vl(\vs)=\frac{\mu}{2}||{\bf H}\vs-\vy||^2,~\vm(\vs)=\lambda\phi({\cal D}\vs).
\label{eqn:sol1}
\end{equation}
The unconstrained optimization of~\ref{eqn:argmin} is {split} and we introduce  two variables $\vs$ and $\vu$ under the constraint that $\vu= {\cal D }\vs$.  As a result, the new problem is given by
\begin{equation}
(\hat{\vs},\hat{\vu})= \displaystyle \argmin{\vu = {\cal D }\vs} \left( \vl(\vs)+\vm(\vu)  \right).
\label{eqn:sol3}
\end{equation}
With the given constraint, it is clear that  $\vu={\cal D }\vs,$ and therefore the solutions of equation~\ref{eqn:argmin} and~\ref{eqn:sol3} have to be the same. This method is favored with respect to the latter unconstrained problem because it decomposes two functions, meaning that we can use different techniques for each one of them. A penalty function is added to~\ref{eqn:sol3}  to penalize the difference between $ {\cal D }\vs$ and $\vu$:
\begin{equation}
(\hat{\vs},\hat{\vu})=  \displaystyle \argmin{\vs,\vu} \left( \vl(\vs)+\vm(\vu)+\frac{\beta}{2}\left \| {\cal D }\vs-\vu\right \| ^{2} \right)
\label{eqn:sol4}.
\end{equation}
The constant parameter $\beta$ controls the magnitude of the penalty function.  It is important to note that the choice of $\beta$ will affect the rate of convergence of the algorithm. As the value of $\beta$ is increased, the constraint of $ {\cal D }\vs=\vu$ is more aggressively enforced. However, as  $\beta$ becomes large, the problem becomes stiffer. The choice of $ \beta$ falls into the so-called \emph{Goldilocks Principle} :
\begin{enumerate}
\item If $\beta$ is too big then the algorithm becomes ill conditioned.
\item  If $\beta$ is too small then the algorithm may converge very slowly.
\end{enumerate}}
\paragraph{}{The proper value of $\beta$ is always a trade-off between constraints 1 and 2.  In this work, we take $\beta$ constant: $\beta =256$.  A thorough study of the variable splitting technique is available in~\citet{Wang2008}. Going back to our particular problem, the splitting technique leads us to consider the constrained problem:
\begin{equation}
\argmin{\vs,\, \vu}  \| {\bf H} \vs-\vy \|^2+  \lambda\phi(\vu) \\
\mbox{s.t. }  \hspace{0.8cm}   \vu={\cal D}\vs,
\label{eqn:argmin2}
\end{equation}
which is then written in the unconstrained form, as explained above:
\begin{equation}
\argmin{\vs, \, \vu} \| {\bf H} \vs-\vy \|^2+ \lambda\phi(\vu)+ \frac{\beta}{2}||{\cal D}\vs-\vu||^2
\label{eqn:opti}.
\end{equation}
The problem formulated in the form of eq.~\ref{eqn:opti} can be solved by an alternating minimization scheme, that is, by solving two subproblems iteratively~\citep{badri:tel-01239958,Ghayem2018}:
\setlength{\abovedisplayskip}{1.5pt}
\setlength{\belowdisplayskip}{2pt}
\begin{equation}
(S1) \ \argmin{\vu} \lambda  \phi(\vu)+ \frac{\beta}{2}||{\cal D}\vs-\vu||^2
\label{eqn:proxy1}
,\end{equation}
\setlength{\abovedisplayskip}{1pt}
\setlength{\belowdisplayskip}{1pt}
\begin{equation}
(S2)  \ \argmin{\vs} ||{\bf H}\vs-\vy||^2+ \frac{\beta}{2}||{\cal D}\vs-\vu||^2
\label{eqn:proxy2}
.\end{equation}
The algorithm decouples the objective function on variables $\vu$ and $\vs$ and minimizes on them independently. This is done with a $\vu$-minimization step (\ref{eqn:proxy1}) and an $\vs$-minimization step  (\ref{eqn:proxy2}). Sub-problem (S2) is quadratic and is implemented efficiently using the Fourier transform as described below. On the other hand, in subproblem (S1), the most interesting priors for deconvolution and/or noise reduction are often neither differentiable nor convex. For this reason, the solution of eq. \ref{eqn:proxy1} is calculated using shrinkage formulae for {proximal operators}, which we briefly describe now (the reader is referred to~\citep{lorenz2007,Rockafellar2009,OPT-003} for a more complete explanation).  }
\paragraph{}{The proximal operator of a proper\footnote{$\psi$ is {\it proper}  if the  set $\{ \vx \in \mathbb{R}^n ~|~\psi(\vx) < +\infty \} $ is not empty.} and lower semicontinuous  function $\psi: \mbox{dom}_{\psi} \subset \mathbb{R}^n \rightarrow ] -\infty, +\infty ]$ is the set-valued function ${\cal P}(\psi) : \mathbb{R}^n \rightarrow 2^{\mathbb{R}^n}$ defined by~\citep{Rockafellar2009}:
\begin{equation}
{\cal P}(\psi) ({\bf x}) = \displaystyle \left\{ \argmin{\vu \in \mbox{dom}_{\psi}} \displaystyle \left( \frac{1}{2} \| \vx - \vu \|_2^2 + \psi(\vu) \right) \right\},
\label{eqn:defproximal}
\end{equation}
where $2^{\mathbb{R}^n}$ is the set of all subsets of ${\mathbb{R}^n}$ (there can be many points $\vu$ that realize a minimum, or no points at all). The {subgradient} of $\psi$ is the subset of $\mathbb{R}^n$ $\partial \psi (\vx) = \{ \vy ~|~\psi({\bf z}) - \psi(\vx) \geq \vy^T\cdot ({\bf z} - \vx) ~\forall \, {\bf z} \in \mbox{dom}_{\psi} \}$. When $\psi$ is differentiable, the subgradient is the usual gradient vector, whereas when $\psi$ is not differentiable, the subgradient may contain a whole continuum of values. For instance if $\psi(\vx) = |\vx|$ ($\vx \in \mathbb{R}$), then $\partial \psi (0) = [-1, +1]$. A point $\vx$ minimizes $\psi$ if and only if $0 \in \partial \psi(\vx)$. When $\psi$ is differentiable, this means $\nabla \psi(\vx) = 0$. If $\psi$ is convex,  $\vx$ is unique. It can be shown in general that ${\cal P}(\lambda \psi) ({\bf x})  \subset (I_{\mathbb{R}^n} + \lambda \partial \psi)^{-1}(\vx)$ with equality when $\psi$ is convex. $(I_{\mathbb{R}^n} + \lambda \partial \psi)^{-1}(\vx)$ is understood to be the set $\{ {\bf z} \, | \, \vx \in  (I_{\mathbb{R}^n} + \lambda \partial \psi)({\bf z})\}$. It is then shown that a minimum of $\psi$ is a fixed point of its proximal operator~\citep{OPT-003}. Consequently, finding a minimum of $\psi$ can be implemented in the form of an iterative process applied on ${\cal P}(\psi)$: ${\bf x}_{k+1} = {\cal P}(\psi)({\bf x}_k)$, even when $\psi$ is neither convex nor differentiable. When $\psi$ is not convex, the iterative scheme may converge towards a local minimum of $\psi$. Then, since the second term $\displaystyle \frac{\beta}{2}||{\cal D}\vs-\vu||^2$ of eq.~\ref{eqn:proxy1} is $\geq 0$, subproblem (S1) is solved using the iteration procedure applied on the proximal operator ${\cal P}(\lambda \phi)$. The potential function $\phi$ and its proximal operator are given in section~\ref{sec:reg}.}
\paragraph{}{For a fixed $\vu$, eq. \ref{eqn:proxy2} is quadratic in $\vs$, and the minimizer $\vs$ is calculated by taking the derivative of eq.  \ref{eqn:proxy2}  and setting it to zero.  We get:
\begin{equation}
({\cal D}^{T}{\cal D}+\frac{2}{\beta}{\bf H}^{T}{\bf H})\vs={\cal D}^{T}\vu+\frac{2}{\beta}{\bf H}^{T}\vy
\label{hessian}
,\end{equation}
with $T$ denoting transposition. After splitting the discrete gradient operator we get:
\begin{equation}
({\cal D}^{T}_{x}{\cal D}_{x}+{\cal D}^{T}_{y}{\cal D}_{y}+\frac{2}{\beta}{\bf H}^{T}{\bf H})\vs={\cal D}^{T}_{x}u_{x}+{\cal D}^{T}_{y}u_{y}+\frac{2}{\beta}{\bf H}^{T}\vy
\label{hessian-split}
.\end{equation}
The Hessian matrix on the left-hand side of eq.~\ref{hessian} can be diagonalized by 2D discrete Fourier transform $\mathcal{F}$. Using the convolution theorem of Fourier transforms, we find that $\vs$ is equal to:
\begin{equation}
\label{eqn:fourier}
\tiny {\mathcal{F}^{-1}\left(\frac{\overline{\mathcal{F}}({\cal D}_{x})\odot \mathcal{F}(u_{x})+\overline{\mathcal{F}}({\cal D}_{y})\odot \mathcal{F}(u_{y})+(\frac{2 }{\beta })\overline{\mathcal{F}}({\bf H})\odot\mathcal{F}(\vy)}{\overline{\mathcal{F}}({\cal D}_{x})\odot \mathcal{F}({\cal D}_{x})+\overline{\mathcal{F}}({\cal D}_{y})\odot\mathcal{F}({\cal D}_{y})+(\frac{2 }{\beta })\overline{\mathcal{F}}({\bf H})\odot \mathcal{F}({\bf H})}\right) }
,\end{equation}
where $\mathcal{F}$ denotes the Fourier transform, “ $\bar{  }$ ” denotes complex conjugation and “$\odot$” component-wise multiplication; the division is also computed component-wise. An advantage of this formulation is that fast-Fourier transform can be used to solve eq.~\ref{eqn:proxy2} to reduce computational complexity. In the following section, we introduce the general regularization function, which is used in this paper for deconvolution.}
\section{A generalized regularization function }
\label{sec:reg}
\paragraph{}{We make use  of  potential functions introduced by~\citep{golami2011}, which are used as $\phi$ in eq.~\ref{eqn:math} and are flexible enough to include in their formulation the classical  $\ell_p$ priors, the latter being those mainly used in sparse image processing. We consider a family of potential functions:
\vspace{0.5cm}
\begin{equation}
\varphi_q^p(\vx)=\left\{\begin{matrix}
\frac{1}{q}(1-(|\vx|^p+1)^{-q })&q\neq0 \\ 
\log(\left |\vx  \right |^p+1)&q=0,
\end{matrix}\right.
\label{eqn:golami}
\end{equation}
where $\vx \in \mathbb{R}$. Eq.~\ref{eqn:golami} defines a family of potential functions indexed by $p$ and $q$, each with its own characteristics.  Some different choices of $p$ and $q$ lead to some known functions as presented in Table ~\ref{tbl:proxy}.
\begin{table}[h]
\caption{Potential functions and their corresponding expression according to the values of $p$ and $q$.}  
\begin{center}
\begin{tabular}{|c|c|c|} 
\hline
\textbf{\textit{p}} & \textbf{\textit{q}} & \textbf{\textit{Expression of $\varphi_q^p(\vx)$}}                                                                                                                                     \\ 
\hline 
p                   & -1                  & $ \left | \vx \right |^{p}$ \\ 
\hline 
1                   & 1                   & $ \frac{\left | \vx \right |}{\left | \vx \right |+1} $  \\ 
\hline 
1                   & 3                   & $ \frac{1}{3}(1-\frac{1}{(\left | \vx \right |+1)^{3}})  $                    \\
\hline
\end{tabular}
\end{center}
\label{tbl:proxy}
\end{table}
Each potential function can be used as prior information in the second term of equation~\ref{eqn:math}. For instance, nonconvex penalties can achieve better performance to promote sparsity~\citep{lorenz2007}. As a standard example, we note that, according to the table shown, the function $\varphi_{-1}^p(\vx) = | \vx |^p$ and the potential function  $\phi(\vu)=||\vu||^p_p= \displaystyle \sum_i  \varphi_{-1}^p(u_i) = \sum_i |u_i|^p$  make use of the $\ell_p$ quasi-norm when  $0< p< 1$). In our case, and referring to eq.~\ref{eqn:proxy1}, we make use of the following potential function:
\begin{equation}
\phi(\vu) = \displaystyle \sum_i \varphi_q^p(u_i).
\label{definition-potential}
\end{equation}
Then, according to theorem (2.1) of~\citep{OPT-003}, one has
\begin{equation}
{\cal P}\left(\phi \right)(\vx) = ({\cal P}\left(\varphi_q^p\right)(x_1), \dots, {\cal P}\left(\varphi_q^p\right)(x_n)),
\label{proximal-potential-function}
\end{equation}
i.e., the computation of the proximal operator ${\cal P}\left(\phi \right)$ can be done component-wise.}
\paragraph*{}{There are some cases where a closed formula of the proximal operator ${\cal P}(\phi)$ can be derived. As an important example, let us consider the case $\varphi_{-1}^p$ (which leads, as seen above, to the nonconvex penalty  $\phi(\vu)=||\vu||^p$ when $0 < p < 1$). Then, according to theorem 3.2 of~\citep{lorenz2007},   we have the following result:
\begin{itemize}
		\item if $1 < p < +\infty $ \begin{equation}  {\cal P}(\varphi_{-1}^p)(\vx) = x + p \mbox{sign}(\vx)|\vx|^{p-1}, \label{shrinkage-lp1}\end{equation}
		\item if $ p= 1$ \begin{equation}  {\cal P}(\varphi_{-1}^p)(\vx) = (|x| - 1)_{+} \mbox{sign}(\vx), \label{shrinkage-lp2}\end{equation}
		\item if $0 < p < 1$,  \begin{equation}  
			{\cal P}(\varphi_{-1}^p)(\vx)=\left\{\begin{matrix}
				\displaystyle 0 & \mbox{if}\, |\vx| \leq \lambda_{\mathrm{eff}}\\ 
				\vx +p|\vx|^{p-1}\mbox{sign}(\vx) &  \mbox{if}\, |\vx| \geq \lambda_{\mathrm{eff}}
			\end{matrix}\right.
			\label{shrinkage-lp3}
			,\end{equation}
\end{itemize}
where $(x)_{+}$ denotes the positive part of $x$ (i.e., $\max(x,0)$) and the threshold value $\displaystyle  \lambda_{\mathrm{eff}} = \frac{2-p}{2 - 2p}(2(1-p))^{\frac{1}{2-p}}$. When $0 < p < 1,$ the function $\varphi_{-1}^p$ is not convex and the proximal operator $ {\cal P}(\varphi_{-1}^p)$ is multivalued at $ \lambda_{\mathrm{eff}}$. However, there is no such closed formula known  for a general function $(\varphi_{q}^p)$. In~\citep{golami2011}, the authors propose the following approximation of $ {\cal P}(\varphi_{q}^p)$, which is valid for $0 < p \leq 2$:
	\setlength{\abovedisplayskip}{2pt}
	\setlength{\belowdisplayskip}{2pt}
\begin{equation}
{\cal Q}\left(\frac{\lambda }{\beta}\varphi_q^p\right)(\vx)=\left\{\begin{matrix}
\displaystyle  \left(1-\frac{\lambda p}{\beta}\left(\frac{|\vx|^{p-2}}{(|\vx|^{p}+1)^{q+1}}\right)\right)_+ \vx &\mbox{if} \  |\vx|>\eta_{\frac{\lambda }{\beta}} \\ 
0& \mbox{otherwise,}
\end{matrix}\right.
\label{eqn:proxy3}
\end{equation}
where ${\cal Q}\left(\frac{\lambda }{\beta}\varphi_q^p\right)$ is the approximated proximal operator of the function $\frac{\lambda }{\beta}\varphi_q^p$ , and the threshold value $\eta_{\frac{\lambda }{\beta}}$  can be computed numerically. As shown in eq.~\ref{eqn:proxy3}, the approximated proximal operator is given by a simple closed-form expression for values bigger than $\eta_{\frac{\lambda}{\beta}}$ and set to zero if smaller.  The proposed general regularization and its approximation induce sparsity in the sense that a number of values, depending on the strength of regularization, will be exactly equal to zero. For a visual demonstration, in Fig.~\ref{fig:proxy} we depict the shrinkage thresholding operators for the case of the $\ell_p$ quasi-norm, where we show a comparison with the exact numerical solution provided in~\citep{lorenz2007} (eq. \ref{shrinkage-lp3}), which is displayed in Fig.~\ref{fig:proxy}. The exact solution is represented by a {solid line} with its corresponding approximation shown as a {dashed line}. 
	We point out that the accuracy of the approximation is very high.}
	\newlength\fheight 
	\newlength\fwidth 
 \begin{figure*}[h]
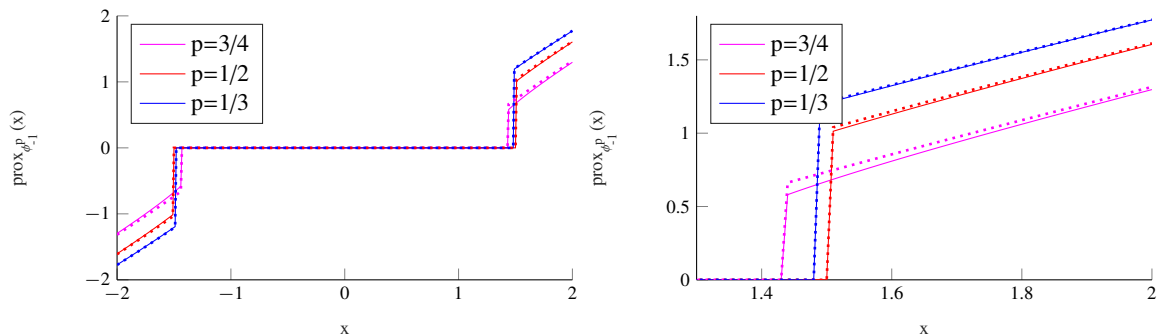

	\scriptsize
	\setlength
	\fheight{3.5cm}
	\setlength\fwidth{6.3cm}
	\definecolor{mycolor1}{rgb}{1.00000,0.00000,1.00000}

\caption{Shrinkage thresholding operators. (Left) Exact and approximate proximal operator of $\varphi_{-1}^p(\vx)=|\vx|^p$ with $\beta=1$.  The approximation is highly accurate in comparison to the exact solution. Only small variations can be observed around the threshold value. (Right) Zoom onto the threshold region.}
\label{fig:proxy}
\end{figure*}
\section{Algorithmic implementation}
\label{sec:algo}
\paragraph{}{	We now give the overall algorithm using the general framework for the subproblem (S1). As outlined in Algorithm~\ref{algo:proxy} below,  we minimize eq.~\ref{eqn:opti} by solving the $\vu$ and $\vs$ subproblems separately until the algorithm {converges}. The question of the choice of parameters is addressed for example in~\citet{Wang2008}; here we keep $\beta$ constant ($\beta= 256$), which makes our algorithm simpler and reduces computational complexity. 
	In step 2 of the algorithm,  the solution of subproblem (S1) of equation~\ref{eqn:proxy1} is calculated using the approximate proximal operator of equation~\ref{eqn:proxy3}.  
	On step 3 the $\vu$ obtained from step 2 is used to calculate the $\vs$ of subproblem (S2) of eq.~~\ref{eqn:proxy1} with eq.~\ref{eqn:proxy2}. This process is repeated until the algorithm is converged.
	\newcommand\eindent{\endgroup}
	\setlength{\abovedisplayskip}{0 pt}
	\setlength{\belowdisplayskip}{0 pt}
	\begin{algorithm}[h]                        
		\SetAlgoLined
		\textbf{Input}: initialize $\vs$, ${\bf H}$, $\lambda >0, \beta, i=0, j=0$\\ 
		1:\hspace{0.3cm}\textbf{while} not converged  \textbf{do}\\
		2:\hspace{0.3cm}Compute $\vu_j$ according to (eq.~\ref{eqn:proxy1}) for fixed $\vs$; \\
		3:\hspace{0.3cm}Compute $\vs_j$ using $\vu_j$ according to (eq.~\ref{eqn:proxy2}) ; \\ 
		4:\hspace{0.3cm}Compute  $\mbox{Cost}(j)$ ; \\ 
		5:\hspace{0.3cm}$j\leftarrow j+1$ \\
		6:\hspace{0.3cm}\textbf{end while}\\
		7: return  $\vs_i$;\\
		8: return $\mbox{Cost}$;
		\caption{\small{Image deconvolution using approximate proximal operators. See text for the definition of the $\mbox{Cost}$ function.}}
		\label{algo:proxy}
\end{algorithm}}
	\paragraph{}{The Cost function is defined in the following way. In Algorithm~\ref{algo:proxy}, for a given $\vy\in \mathbb{R}^n$, ${\bf H} \in \mathbb{R}^{n\times n}
		$ and recovered $\hat{\vs}\in \mathbb{R}^n$ at iteration $j$, the Cost is defined as : $
		\displaystyle
		\mbox{Cost}(j)=||{\bf H}\hat{\vs_{j}}-\vy||^2+\lambda\phi(D\hat{\vs_{j}})$. At iteration $j+1$, $\mbox{Cost}(j+1)$ is computed and compared with the cost at iteration $j$ using a threshold value $\epsilon$: if $|\mbox{Cost}(j+1)-\mbox{Cost}(j) |\leqslant \epsilon \;\displaystyle \mbox{then}$ the algorithm  continues; else $\hat{\vs_{j}}$ is returned as the result. We note that this definition uses the problem itself as a means to measure how close and effective the $\hat{\vs}$ is to the real $\vs$ that we are trying to recover.}
	\section{Choice of the potential function and the values of parameters}
	\label{sec:choice}
\begin{figure*}[h]
	\centering
	\includegraphics[scale=0.35]{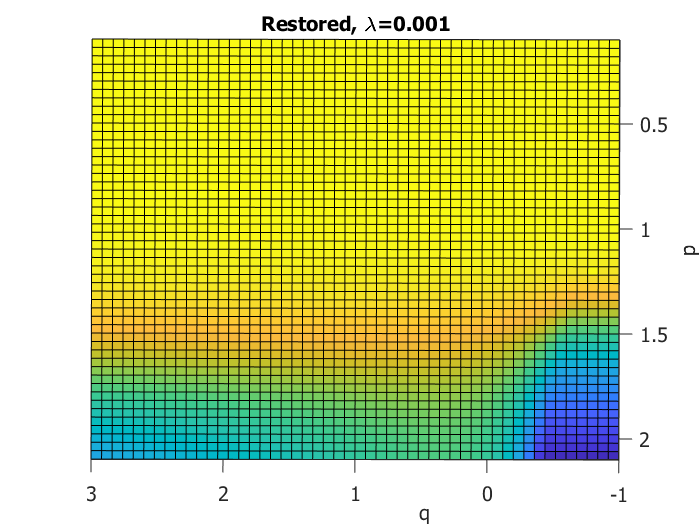} \includegraphics[scale=0.35]{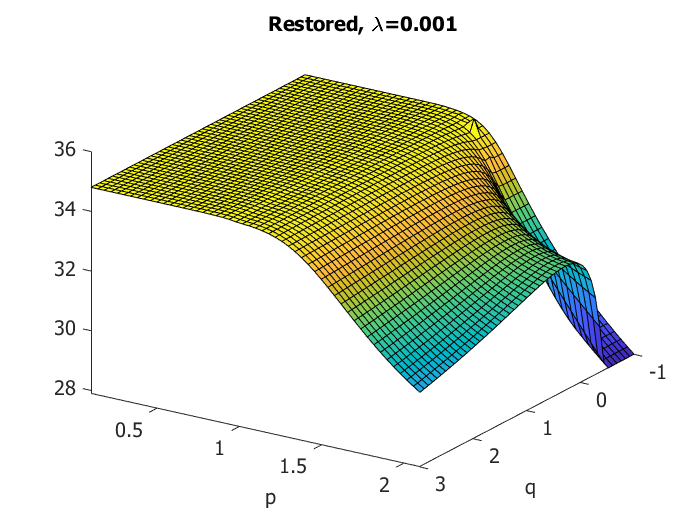}
	\caption{Surface of PSNR values as a function of $p$ and $q$ after deconvolution using the regularizer $\varphi_q^p(\vx)$. The data file is a  MHD simulation output as described in section~\ref{sec:deconvolution}.  In the  experiment, the beam effect  ${\bf H}$ is the beam kernel defined in section~\ref{sec:deconvolution}. The lambda parameter is $\lambda = 10^{-3}$. For a discrete set of  $(p,q)$ values in the range $[0,2] \times [-1, 3]$ (50 equally spaced points on each axis)), we draw a color point representing the PSNR between the original (with no beam) image and the reconstructed one after applying the optimization process eq.~\ref{eqn:math}. The higher the PSNR, the better the reconstruction. Left and right panels show respectively a 2D (orthographic) top view and a 3D view of the PSNR surface.}
	\label{fig:testpq0}
\end{figure*}
\begin{figure*}[h]
	\centering
	\includegraphics[scale=0.35]{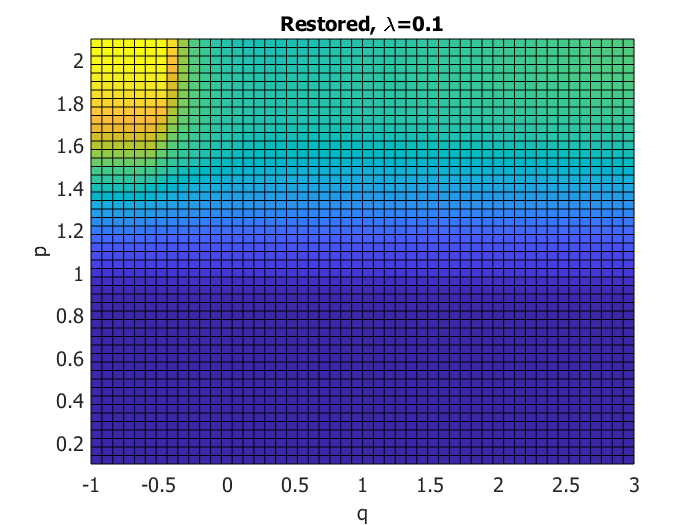} \includegraphics[scale=0.35]{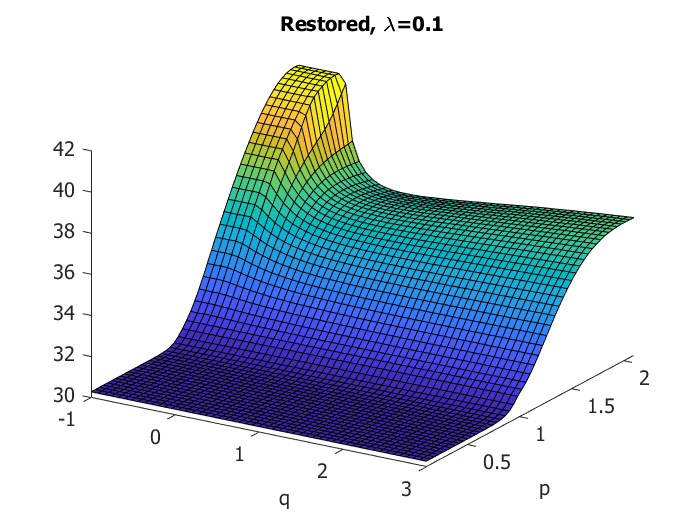}
	\caption{Same experiment as in Fig.~\ref{fig:testpq0}, but with $\lambda = 10^{-1}$. This figure demonstrates the sensitivity of the optimization process in Eq.~\ref{eqn:math} with respect to parameter $\lambda$.}
	\label{fig:testpq1}
\end{figure*}
\paragraph*{}{
We address in this section the problem of determining the values of the parameters $p$, $q,$ and $\lambda,$ which must be injected into the minimization process~\ref{eqn:opti} when the prior is given by the definition~\ref{eqn:golami} and when this minimization process is applied to an actual observation map. One of the main difficulties in processing astronomical signals is the lack of a ground truth, which is otherwise used to determine the parameter values: in addition, in our case, the difficulty is aggravated by the fact that we have only one {\it Herschel} acquisition for any given cloud. This prevents any supervised validation on a large number of acquisitions. And simulation model outputs are available.}
\paragraph*{}{
We therefore used the simulation data ---for which we have both the raw data (without beam or noise) and this same data on which beam plus noise has been added--- to determine the values of the parameters. Hence we took a sample from the MHD simulation data described in section~\ref{sec:deconvolution} (data file NH1\-d1cf05bx10rms36000\_vz ) and convolved the image with $\bf H$, where ${\bf H}$ is the beam kernel defined in section~\ref{sec:deconvolution}. Regarding the noise,  we used one $1024\times 1024$ sample with added Gaussian noise ($\sigma=10^{-2}$).}
\paragraph*{}{
Next, we used our algorithm with various potential function configurations (values of $p$ and $q$ of $\varphi_q^p$) to obtain different results based on the corresponding regularizer. The initial values of $p$ and $q$ correspond to $q \in [-1,3]$ and $p \in [0,2],$ with each range divided into 50 equally spaced points. Although we carried out experiments with value intervals for $p$ and $q$, in practice we only consider priors in the form of an $l^p$ norm, which imposes $q=-1$ according to the theory presented in Appendix \ref{sec:reg}. The cost surfaces shown in figures \ref{fig:testpq0} and ~\ref{fig:testpq1} must therefore be read in our context at $q=-1,$ but we show the results for all values of $q$.}
\paragraph*{}{
Regarding $\lambda$, we note that this parameter defines the weight of the prior in equations~\ref{eqn:argmin2} and~\ref{eqn:opti} and so controls the sparsity of the result. As we are interested in enhancing the filamentary structures in interstellar clouds, which typically have sparse gradients, $\lambda$ should not be excessively small. Here we use  two values, $\lambda = 10^{-3}$ and $\lambda = 10^{-1}$ . For each value $(p,q) \in [0,2] \times [-1,3]$, we solve the problem~\ref{eqn:opti} and obtain a signal for which we can compute the PSNR with the original unaltered data. Figures~\ref{fig:testpq0} and ~\ref{fig:testpq1} show the PSNR surface corresponding to the two values $\lambda = 10^{-3}$ and $\lambda = 10^{-1}$.  The resulting PSNR surfaces shown in figures~\ref{fig:testpq0} and ~\ref{fig:testpq1}, calculated on the simulation output, show a clear dependence of the result on $\lambda$, but those at $\lambda = 0.1$ have a higher PSNR, while ensuring good sparsity. We therefore set $\lambda = 0.1$.}
\paragraph*{}{
In the end, parameter $p$ must be adjusted. Our experiments show that values close to $p=2$ in the range $[1.5, 2]$ provide good PSNRs on a simulation output. However, on real data, we find that smaller values of $p$ ---typically between 1.3 and 1.7--- give images where the filaments are better contrasted. This is not surprising because the resolution of the simulations prevents us from obtaining filaments on the scale of the observational data. In addition, we must not forget that it is a noise reduction model that is proposed, with probably very
simplistic assumptions about this noise  compared to reality: the Gaussian noise used to simulate the noise corresponds to an approximation of the real noise in the data, which is multimodal and more complex. This is why, in the processing of real observation maps, the result must be visually evaluated by an astronomer, in particular in regards to the thickness and the general appearance of the filamentary structures obtained. We note that on the real data we almost always took $p=1.5,$ except only for Spider, which is mainly made up of very fine filaments for which we had to push to $p=1.3$ to obtain an acceptable rendering. The choice $p=1.6$ for certain ROIS in Taurus is very close to $p=1.5$ and simply offers a slight improvement in the contrast of the filaments. This does not change the spectral statistics, which are almost the same at $p=1.5$ or $p=1.6$.}
\paragraph*{}{
The parameter values we find to be generally valid for {\it Herchel} observations are therefore $\lambda =0.1$, $q=-1$, and $p=1.5$. Some fits are more precise at $p=1.3,  $ while others are more precise at $p=1.5;$ but this does not change the statistics and conclusions that we draw from them. On the other hand, in some observation maps, images of the filaments are sometimes  better rendered by adjusting the value of $p$ to around $p=1.5$.}    
\section{Comparison with the classical gradient}
\label{sec:comparison-gradient}
\begin{figure}[h]
\centering
\includegraphics[width=0.5\textwidth]{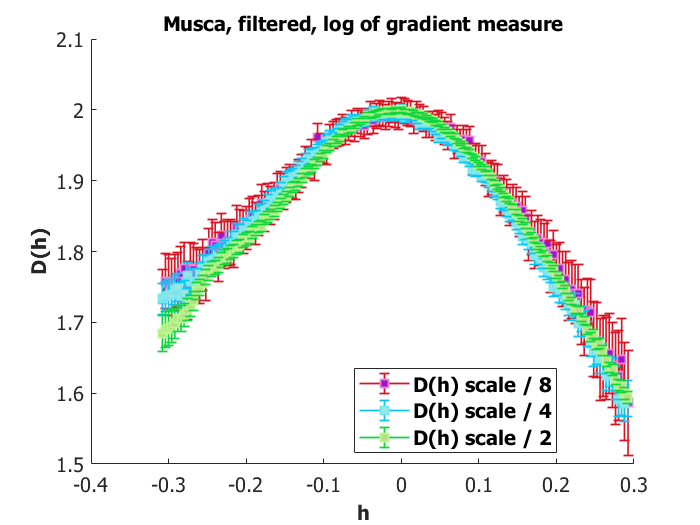} 
\caption{The gradient's norm measure of eq.~\ref{gradient-se} is multiscale. Singularity spectra computed from the gradient norm measure as defined in eq.~\ref{gradient-se} over three consecutively downscaled versions of the Musca observation map (filtered with parameters $p=1.5$, $q=-1$, $\lambda=0.1$) using a reverse bi-orthogonal discrete wavelet transform of order 4.4. The three spectra coincide, which means that the gradient's norm measure is multiscale.}	
\label{gradient-measure-multiscale}
\end{figure}

\paragraph*{}{As the singularity exponents display the  transitions within a signal, one might question whether the filamentary coherent structures shown could equally be rendered by the gradient norms of the Musca map.  It turns out that the gradient norms themselves  possess a high dynamical range  in such a  way that the logarithm of the gradient norms must be considered instead: if ${\bf s}$ is the signal of an observational map, $\log(\| \nabla {\bf s} \| ({\bf x }))$  can indeed be considered as a simple local correlation measure and consequently, an approximation of a singularity exponent at ${\bf x} = (x,y)$. In practice, we compute the gradient of the observation map ${\bf s}$ at ${\bf x}= (x,y)$, $\displaystyle \nabla {\bf s}  ({\bf x}) = \left( \displaystyle \frac{\partial {\bf s} }{\partial x} ({\bf x}),  \displaystyle \frac{\partial {\bf s} }{\partial y} ({\bf x})\right)$ in Fourier space:
\begin{equation}
-{\bf i}\, {\bf x} \odot \nabla {\bf s}({\bf x}) = {\cal F}^{-1} \displaystyle \left( {\bf i} \, {\bf f} \odot {\cal F}({\bf s})({\bf f})\right),
\label{gradient-Fourier}
\end{equation}
where ${\bf f}=(f_1, f_2)$ is the frequency vector, ${\bf x} =(x,y)$ denotes spatial coordinates, ${\cal F}$ is the Fourier transform, ${\cal F}^{-1}$ the inverse Fourier transform, $\odot$ means component-wise multiplication, and ${\bf i}$ is the imaginary unit. The $\log$ of the gradient's norm is defined as:
\begin{equation}
\label{gradient-se}
{\bf h}({\bf x}) = 
\left\{ \begin{matrix}
\displaystyle  \log \left( \displaystyle  \frac{\| \nabla {\bf s}({\bf x}) \| }{\langle \| \nabla {\bf s}\| \rangle} \right)/{\log ({\bf l})} & \mbox{if } \displaystyle   \displaystyle  \frac{\| \nabla {\bf s}({\bf x}) \|}{\langle \| \nabla {\bf s}\| \rangle} > \varepsilon \\
\displaystyle \frac{\log(\varepsilon)}{\log(\bf{l})} & \mbox{else,}
\end{matrix}\right\} ,
\end{equation}
where ${\bf l} = (l_x l_y)^{-\frac{1}{2}}$, $l_x$, $l_y$ are the $x$ and $y$ lengths in pixel units of the array of values associated to the observational map ${\bf s}$, ${\langle \| \nabla {\bf s}\| \rangle} $ is the average of the gradient norms over the spatial domain of the observation map, and $\varepsilon$ is a threshold value, chosen here to be $\varepsilon = 10^{-30}$ (lower threshold used to cut null values). The ${\bf h}({\vx})$ defined by eq.~\ref{gradient-se} are the singularity exponents of a local correlation measure defined with a gradient vector obtained from the Fourier transform. This correlation measure truly has multiscale properties, as shown in figure~\ref{gradient-measure-multiscale}, in which we display the singularity spectra computed from that gradient norm measure over three consecutively downscaled versions of the Musca observation map using a reverse bi-orthogonal discrete wavelet transform of order 4.4. The three spectra coincide, which means that the  ${\bf h}({\vx})$  defined by eq.~\ref{gradient-se} exist in the limit.}
\begin{figure}[h]
	\centering
	\includegraphics[width=0.44\textwidth]{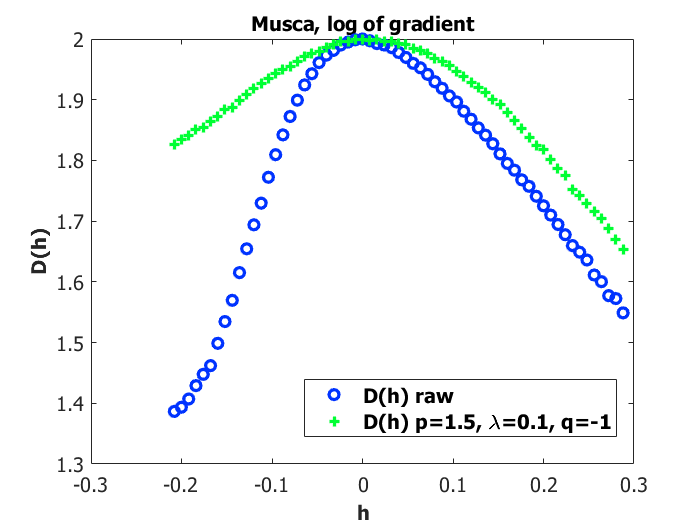} 
	\caption{ Comparison between the singularity spectrum of Musca unfiltered (blue) computed with ${\bf h}({\bf x})$ given by eq.~\ref{gradient-se} and the singularity spectrum of Musca with noise reduction, with $p=1.5$, $q=-1$, $\lambda=0.1,$ computed with ${\bf h}({\bf x})$ given by eq.~\ref{gradient-se}.}
	\label{musca-se-GDzoom2}
\end{figure}
\begin{figure}[h]
	\centering
	\includegraphics[width=0.5\textwidth]{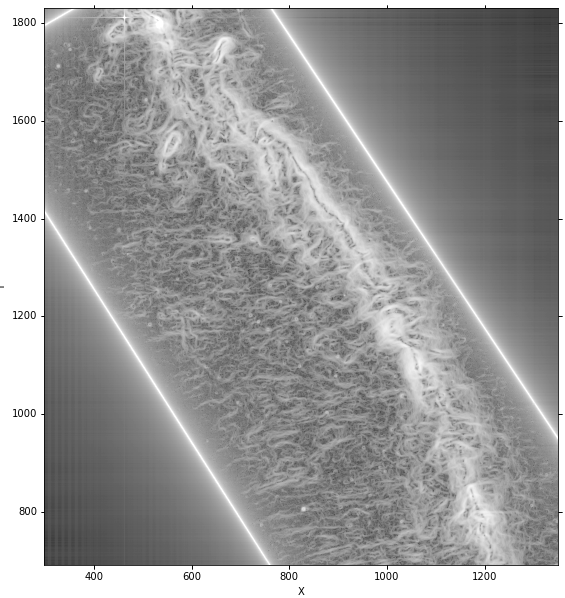} 
	\caption{ Visualization of the $\log$ of gradient norms on the {\sl Herschel} observation map of Musca, after noise reduction with $p=1.5$, $q=-1$, $\lambda=0.1$. Although the $\log$ of gradient norms enhances the striations, the resulting spectrum, shown in figure~\ref{musca-se-GDzoom2}, is very different from the one obtained with the correlation measure.}
	\label{musca-se-GDzoom1}
\end{figure}
\paragraph*{}{Figure~\ref{musca-se-GDzoom1} displays the resulting gradient norms, as defined by eq.~\ref{gradient-se}.  The resulting striations and coherent filamentary structures closely resemble those of figure~\ref{musca-se-zoomb}, but these values do not encode the  same statistics of turbulence, as shown in figure~\ref{musca-se-GDzoom2}:we display two singularity spectra computed with ${\bf h}({\bf x})$  from eq. ~\ref{gradient-se}; the blue curve is the spectrum of the raw map and the green curve the spectrum of the filtered one (with the same parameter values). The two curves are extremely dissimilar, and, most importantly, the spectrum of the filtered map is above the spectrum of the raw map in all parts, which is a strange and unexpected result: this suggests there are more low-dimensional structures (which are therefore typically filamentary) in the raw data than in the filtered data. This is contradictory even by simple visual inspection of the data. Consequently, although the $\log$ of gradient norms reveals interesting filamentary structures on observation maps with noise reduction, we must keep the values of the singularity exponents given by formula (B.7) in~\cite{yahia2021} for both visualization and the computation of correct statistics of turbulence. This  was carried out systematically for the other interstellar clouds in the study  presented in the main text.}
\end{appendix}
\end{document}